\documentclass[12pt,preprint]{aastex}

\received{2 November 2013}
\accepted{12 May 2014}

\shorttitle{Extremely young objects in B1}
\shortauthors{Hirano \& Liu}

\begin{document}

\title{Two Extreme Young Objects in Barnard 1-b}

\author{NAOMI HIRANO\altaffilmark{1,2} and FANG-CHUN LIU\altaffilmark{3,4}}

\altaffiltext{1}{Academia Sinica,
Institute of Astronomy \& Astrophysics, 11F of Astronomy-Mathematics Building, National Taiwan University, No 1, Sec. 4, Roosevelt Rd, Taipei, 10617, Taiwan, R.O.C.}
\altaffiltext{2}{e-mail: hirano@asiaa.sinica.edu.tw}
\altaffiltext{3}{Max Planck Institut f\"ur Radioastronomie, Auf dem H\"ugel 69, 53121 Bonn, Germany}
\altaffiltext{4}{Member of the International Max Planck Research School (IMPRS) for Astronomy and Astrophysics at the Universities of Bonn and Cologne.}

\begin{abstract}
Two submm/mm sources in the Barnard 1b (B1-b) core, B1-bN and B1-bS, have been studied in dust continuum, H$^{13}$CO$^+$ $J$=1--0, CO $J$=2--1, $^{13}$CO $J$=2--1, and C$^{18}$O $J$=2--1.
The spectral energy distributions of these sources from the mid-IR to 7 mm are characterized by very cold temperatures of $T_{\rm dust} <$ 20 K and low bolometric luminosities of 0.15--0.31 $L_{\odot}$.
The internal luminosities of B1-bN and B1-bS are estimated to be $<$0.01--0.03 $L_{\odot}$ and $\sim$0.1--0.2 $L_{\odot}$, respectively.
Millimeter interferometric observations have shown that these sources have already formed central compact objects of $\sim$100 AU sizes.
Both B1-bN and B1-bS are driving the CO outflows with low characteristic velocities of $\sim$2--4 km s$^{-1}$.
The fractional abundance of H$^{13}$CO$^+$ at the positions of B1-bN and B1-bS is lower than the canonical value by a factor of 4--8.
This implies that significant fraction of CO is depleted onto dust grains in dense gas surrounding these sources.
The observed physical and chemical properties suggest that B1-bN and B1-bS are in the earlier evolutionary stage than most of the known Class 0 protostars.
Especially, the properties of B1-bN agree with those of the first hydrostatic core predicted by the MHD simulations.
The CO outflow was also detected in the mid-IR source located at $\sim$15\arcsec from B1-bS.
Since the dust continuum emission was not detected in this source, the circumstellar material surrounding this source is less than 0.01 $M_{\odot}$.
It is likely that the envelope of this source was dissipated by the outflow from the protostar that is located to the southwest of B1-b.
\end{abstract}


\section{INTRODUCTION}

Probing the initial condition of protostellar collapse is essential to understand the formation process of low-mass stars.
In the earliest phase of star formation, a hydrostatic equilibrium object, so-called  ^^ ^^ first core", is expected to be formed at the center of the collapsing cloud \citep[e.g.][]{Lar69}.
The formation, evolution, and structure of first cores are extensively studied by numerical simulations \citep[e.g.][]{Masu00,Saigo06,Sai08,Comm12}.
Recent studies have shown that the first core formed in a rotating core survives longer (more than 1000 yr) than that of the spherical symmetric case ($\sim$100 yr), and becomes the origin of the circumstellar disk \citep[e.g.][]{Saigo06,Sai08,Machi11}.
Observationally, several first core candidates have been discovered on the basis of high-sensitivity mid-IR observations with the {\it Spitzer} Space Telescope and sensitive mm and sub-mm observations with the Submillimeter Array (SMA);
Cha-MMS \citep{Bel06,Bel11,Tsi13}, L1448-IRS2E \citep{Chen10}, Per-Bolo 58 \citep{Eno10,Dun11}, L1451-mm \citep{Pin11}, CB17 MMS \citep{Chen12}.
Although these candidate sources have not yet been confirmed as first cores, they are likely to be in the earlier evolutionary stage than previously known class 0 protostars.
Therefore, detailed structures of the envelopes and central compact objects of such sources would provide us with clues to understand how the initial object is formed in the starless core.

A dense molecular cloud core Barnard 1-b (B1-b) is one of the most suitable regions to study the earliest stage of protostellar evolution.
B1-b is the most prominent core in the Barnard 1 (B1) star-forming cloud \citep{Wala05, Eno06}, which is one of the highest column density regions in the Perseus cloud complex.
B1-b harbors  two submillimeter sources, B1-bN amd B1-bS,  separated by $\sim$20\arcsec \citep{Hira99}.
These two sources have no counterpart in the 4 bands of IRAC (3.6, 4.5, 5.8, and 8.0 $\mu$m) and even in the 24 and 70 $\mu$m bands of MIPS observed by the {\it Spitzer} legacy project ^^ ^^ From Molecular Cores to Planet Forming Disks" \citep[c2d;][]{Evans03}.
Although there is an infrared source detected by {\it Spitzer} space telescope \citep{Jor06}, the location of this infrared source does not coincide with neither of the submillimeter sources.
These imply that the two submillimeter sources in B1-b are colder than most of the class 0 protostars, which are usually detected in the Spitzer wavebands.
Recently, \citet{Pez12} have detected the far-infrared emission from B1-bN and B1-bS, and proposed that these sources are candidates for first cores.

The distance to B1 is estimated to be 230 pc by \citet{Cern03} from the extinction study. 
On the basis of the parallax measurements of the H$_2$O masers, the consistent distances of 235$\pm$18 pc and 232$\pm$18 pc have been derived for NGC 1333 SVS13 and L1448C, respectively, which are located at $\sim$1$^{\circ}$ west of B1\citep{Hiro08,Hiro11}.
In this paper, we adopt 230 pc as a distance to B1.

In this paper, we report the millimeter and submillimeter observations of B1-b.
We have obtained the high resolution continuum images of these two sources with the Very Large Array (VLA), Nobeyama Millimeter Array (NMA), and the SMA\footnote{The Submillimeter Array (SMA) is a joint project between the Smithsonian Astrophysical Observatory and the Academia Sinica Institute of Astronomy and Astrophysics, and is funded by the Smithsonian Institution and the Academia Sinica.}.
In addition, molecular line maps were obtained with the 45m telescope of Nobeyama Radio Observatory (NRO)\footnote{Nobeyama Radio Observatory (NRO) is a branch of the National Astronomical Observatory, an inter-university research institute operated by the Ministry of Education, Culture, Sports, Science and Technology,  Japan.}, NMA, and the SMA.

\section{OBSERVATIONS AND DATA REDUCTION}

\subsection{Single-dish observations}

\subsubsection{Continuum observations}

The 350 $\mu$m continuum observations were carried out in 1997 February 13 using the Submillimeter High Angular Resolution Camera (SHARC), a 20-pixel linear bolometer array camera, on the  Caltech Submillimeter Observatory (CSO)\footnote{The Caltech Submillimeter Observatory is operated by the California Institute of Technology under cooperative agreement with the U.S. National Science Foundation (AST-0838261)} 10.4 m telescope.
The beam size determined from the map of Mars was 11\farcs3.
Pointing was checked by observing Saturn and Mars, and found to be drifted by $\sim$13$''$, mostly in the azimuth direction, during the observations.  
The zenith opacity at $\lambda$=1.3 mm, monitored by the radiometer at the CSO, was $\sim$0.05, and hence the 350 $\mu$m atmospheric transmission at zenith was $\sim$40\%.
We used the on-the-fly mapping mode in azimuth-elevation grid.
The mapped area was $\sim$2$'$ $\times$ 50$''$ in right ascension and declination.
The rms noise level of the 350 $\mu$m map was $\sim$0.2 Jy beam$^{-1}$.

The 850 $\mu$m observations were made in 1997 August using the Submillimeter Common-User Bolometer Array (SCUBA) at the James Clerk Maxwell Telescope (JCMT)\footnote{The James Clerk Maxwell Telescope is operated by the Joint Astronomy Centre on behalf of the Science and Technology Facilities Council of the United Kingdom and the National Research Council of Canada.}.
Since the array field of view had a diameter of 2.3$'$, we observed one field centered at $\alpha$(2000) = 3$^{\rm h}$33$^{\rm m}$18.84$^{\rm s}$, $\delta$(2000) = 31$^{\circ}$ 07$'$ 33\farcs9.
To sample the field of view, the secondary mirror was moved in a 64-point jiggle pattern. Telescope pointing was checked by observing Saturn.
The absolute flux scale was determined by observing Uranus.
The atmospheric transmission at zenith derived from the sky-dip measurements was $\sim$75\% at 850 $\mu$m.
The beam sizes measured by using the Uranus images was 15$''$.
The rms noise level of the 850 $\mu$m map was $\sim$20 mJy beam$^{-1}$.

\subsubsection{H$^{13}$CO$^+$ and HC$^{18}$O$^+$ $J$=1--0}

The H$^{13}$CO$^+$ $J$=1--0 observations were carried out with the NRO 45 m telescope in 2005 January.
We used the 25 BEam Array Receiver System (BEARS). 
The half-power beam width (HPBW) and the main-beam efficiency of the telescope at 87 GHz were 19$''$ and 0.5, respectively.
The backend was an auto correlator with a frequency resolution of 37.8 kHz, which corresponds to a velocity resolution of 0.13 km s$^{-1}$.
An area of $\sim$4$'\times$4$'$ in right ascention and declination centered at $\alpha$(2000) = 3$^{\rm h}$33$^{\rm m}$21.14$^{\rm s}$, $\delta$(2000) = 31$^{\circ}$ 07$'$ 23\farcs8 was mapped with a grid spacing of 10.25\arcsec, which corresponds to the quarter of the beam spacing (41\farcs1) of BEARS.
The typical rms noise per channel was $\sim$0.4 K in $T_{\rm mb}$.
In order to calibrate the intensity scale of the spectra obtained with BEARS which is operated in DSB mode, we obtained the H$^{13}$CO$^+$ spectra toward the reference center using two SIS receivers operated in SSB mode.
The telescope pointing was checked by observing the nearby SiO maser source toward NML Tau at 43 GHz.

Additional H$^{13}$CO$^+$ and HC$^{18}$O$^+$ $J$=1--0 spectra were obtained with the NRO 45m telescope in 2009 February.
Two isotopic lines were observed simultaneously using the 2SB SIS receiver, T100H/V at the positions of B1-bN and B1-bS. 
The main beam efficiency was 0.43, and the HPBW was 18.4$''$.
The backend was a bank of acousto-optical spectrometers with a frequency resolution of 37 kHz, which corresponds to 0.13 km s$^{-1}$ at 86 GHz.

\subsection{Millimeter-interferometer observations}

The interferometric data were obtained with the VLA, NMA, and the SMA.
The continuum and line observations were summarized in Tables 1 and 2, respectively.

\subsubsection{7 mm observations with the VLA}

The 7 mm data were retrieved from the VLA data archive.
The data were taken in 2004 February using the BnC configuration.
The standard VLA continuum frequency setup (43.3149 and 43.3649 GHz) was used.
The bandwidth was 50 MHz in each sidebnad.
The absolute flux density was set using the observations of 3C48.
The phase and amplitude were calibrated by observing 03365+32185 every 2 minutes.
The data were calibrated and imaged using the Astronomical Image Processing System (AIPS).
The synthesized beam of the natural weighting map was 0\farcs51${\times}$0\farcs45 with a position angle of -71.1$^{\circ}$.
The rms noise level was $\sim$0.11 mJy beam$^{-1}$.

\subsubsection{3 mm observations with the NMA}

Observations of the H$^{13}$CO$^+$ J=1-0 line and the 3 mm continuum emission were made with the NMA from 1998 January to May in three different array configurations.
The primary-beam sizes in HPBW were 86$''$ at 3.5 mm and 78$''$ at 
3.0 mm.
The coordinates of the field center were 
$\alpha$(2000) = 3$^{\rm h}$33$^{\rm m}$21.14$^{\rm s}$, 
$\delta$(2000) = 31$^{\circ} $07$'$ 23\farcs8.
The phase and amplitude were calibrated by observing the nearby quasar IRAS 0234+385.
The flux of IRAS 0234+385 at the time was determined from observations of Mars.
Observations of 3C454.3 and 3C273 were used for bandpass calibrations.

The H$^{13}$CO$^+$ line data were recorded in a 1024 channel digital FFT 
spectrocorrelator (FX) with a bandwidth of 32 MHz and a frequency 
resolution of 31.25 kHz.
The visibility data were calibrated using UVPROC2 software package developed at NRO and Fourier transformed and CLEANed using the MIRIAD package.
In making maps of H$^{13}$CO$^+$, we applied a 45 k$\lambda$ taper and 
natural weight to the visibility data.
The synthesized beam had a size of 7\farcs2${\times}$6\farcs1 with a position angle of -22.7$^{\circ}$.
The rms noise level of the H$^{13}$CO$^+$ maps at a velocity
resolution of 0.25 km s$^{-1}$ was 0.15 Jy beam$^{-1}$,
which corresponds to 0.56 K in $T_{\rm B}$.

The lower and upper sideband continuum signals were recorded 
separately in two ultra-wide-band correlators of 1 GHz bandwidth.
To improve the signal to noise ratio, we combined the data of the
upper and lower sidebands after we checked for consistency.
In making the continuum map, we adopted robust weighting of 1 without UV taper.
The synthesized beam was 3.3$''{\times}$2.9$''$ with a position angle of -73.2$^{\circ}$.
The rms noise level of the continuum map was $\sim$1.4 mJy beam$^{-1}$.

\subsubsection{1.3 mm observations with the SMA}

The 1.3 mm (230 GHz) observations were carried out on September 10 and 11, 2007 with the SMA  \citep{Ho04}. 
The CO $J$=2--1, $^{13}$CO $J$=2--1, C$^{18}$O $J$=2--1, SO $J_K$=5$_6$--4$_5$, and N$_2$D$^+$ $J$=3--2 lines were observed simultaneously with 1.3 mm continuum using the 230 GHz receivers. 
We used seven antennas in the subcompact configuration. 
The primary-beam size (HPBW) of the 6\,m diameter antennas at 1.3 mm was measured to be $\sim$54$''$. 
The phase tracking center of the 1.3 mm observations was at $\alpha$(J2000) = 3$^h$33$^m$21.14$^s$, $\delta$(J2000)=31$^{\circ}$07$'$35\farcs3.
The spectral correlator covers 2 GHz bandwidth in each of the two sidebands separated by 10 GHz. 
Each band is divided into 24 ^^ ^^ chunks'' of 104 MHz width.
We used a hybrid resolution mode with 1024 channels per chunk (101.6 kHz resolution) for the N$_2$D$^+$ $J$=3--2 and C$^{18}$O $J$=2--1 chunks (s23 of the upper and lower sideband, respectively) and 256 channels per chunk (406.25 kHz) for the CO, $^{13}$CO, and SO. 
In this paper, we present the results in continuum, CO, $^{13}$CO, and C$^{18}$O.
The results in N$_2$D$^+$ were presented in \citet{Huang13}.

The visibility data were calibrated using the MIR software package.
The absolute flux density scale was determined from observations of Ceres and Uranus for the data obtained on September 10 and 11, respectively.
A pair of nearby compact radio sources 3C84 and 3C111 were used to calibrate relative amplitude and phase.
We used 3C111 to calibrate the bandpass for the September 10 data except the chunks including CO $J$=2--1 and $^{13}$CO $J$=2--1. 
For two chunks including CO and $^{13}$CO, 3C111 could not be used as a bandpass calibrator because of strong absorption due to the foreground Galactic molecular cloud.
Therefore, we used 3C454.3 and 3C84 to calibrate the bandpass for the CO and $^{13}$CO chunks.
The bandpass of the September 11 data was calibrated by using Uranus.

The calibrated visibility data were Fourier transformed and CLEANed using MIRIAD package.
We used natural weighting to map the data.
The synthesized beam size and the rms noise level of each line are summarized in Table 2.

The continuum map was obtained by averaging the line-free chunks of both sidebands.
To improve the signal to noise ratio, the data of the upper and lower sidebands were combined.
The synthesized beam of the natural weighting map had a size of 6\farcs3$\times$ 4\farcs0 with a position angle of 49$^{\circ}$.
The rms noise level of the 1.3 mm continuum map was $\sim$2.3 mJy beam$^{-1}$ .

\subsubsection{1.1 mm observations with the SMA}

The 1.1 mm (279 GHz) observations were carried out on September 3, 2008 with the SMA.
The 1.1 mm continuum emission was observed simultaneously with the N$_2$H$^+$ $J$=3--2 line using the 345 GHz receivers. 
We used eight antennas in a subcompact configuration. 
The primary-beam size (HPBW) of the 6\,m diameter antennas at 1.1 mm was $\sim$45$''$. 
The phase tracking center was same as that of the 1.3 mm observations.
The spectral correlator covers 2 GHz bandwidth in each of the two sidebands separated by 10 GHz centered at 283.8 GHz. 

The visibility data were calibrated using the MIR software package.
Uranus was used to calibrate the bandpass and flux. 
A pair of nearby compact radio sources 3C84 and 3C111 were used to calibrate the amplitude and phase.
The calibrated visibility data were Fourier transformed and CLEANed using MIRIAD 
package. 
The continuum map was obtained by averaging the line-free chunks.
The natural weighting provided the synthesized beam of 3\farcs3$\times$2\farcs9 with a position angle of 56.8$^{\circ}$.
The rms noise level of the 1.1 mm continuum map was $\sim$3.1  mJy beam$^{-1}$ after combining the upper and lower sidebands.
In this paper, we only present the results in continuum.
The results in N$_2$H$^+$ line were presented in \citet{Huang13}.

\subsection{Combining Single-dish and interferometer data}
In order to fill the short-spacing information that was not sampled by the interferometer,  the H$^{13}$CO$^+$ data obtained with the 45m telescope and the NMA were combined using MIRIAD.
We followed the procedure described in \citet{Tak03}, which is based on the description of combining single-dish and interferometric data by \citet{Vog84}.
The combined H$^{13}$CO$^+$ maps were made with a 45 k$\lambda$ taper and natural weighting, and deconvolved using the MIRIAD task CLEAN.
The synthesized beam size of the combined H$^{13}$CO$^+$ map was 7\farcs7$\times$6\farcs5 with a position angle of $-$22.5$^{\circ}$.
The rms noise level of the combined map at a velocity
resolution of 0.25 km s$^{-1}$ was 0.14 Jy beam$^{-1}$,
which corresponds to 0.46 K in $T_{\rm B}$.

\section{RESULTS}
\subsection{Dense gas distribution in B1}

\begin{figure} 
\epsscale{1}
\plotone{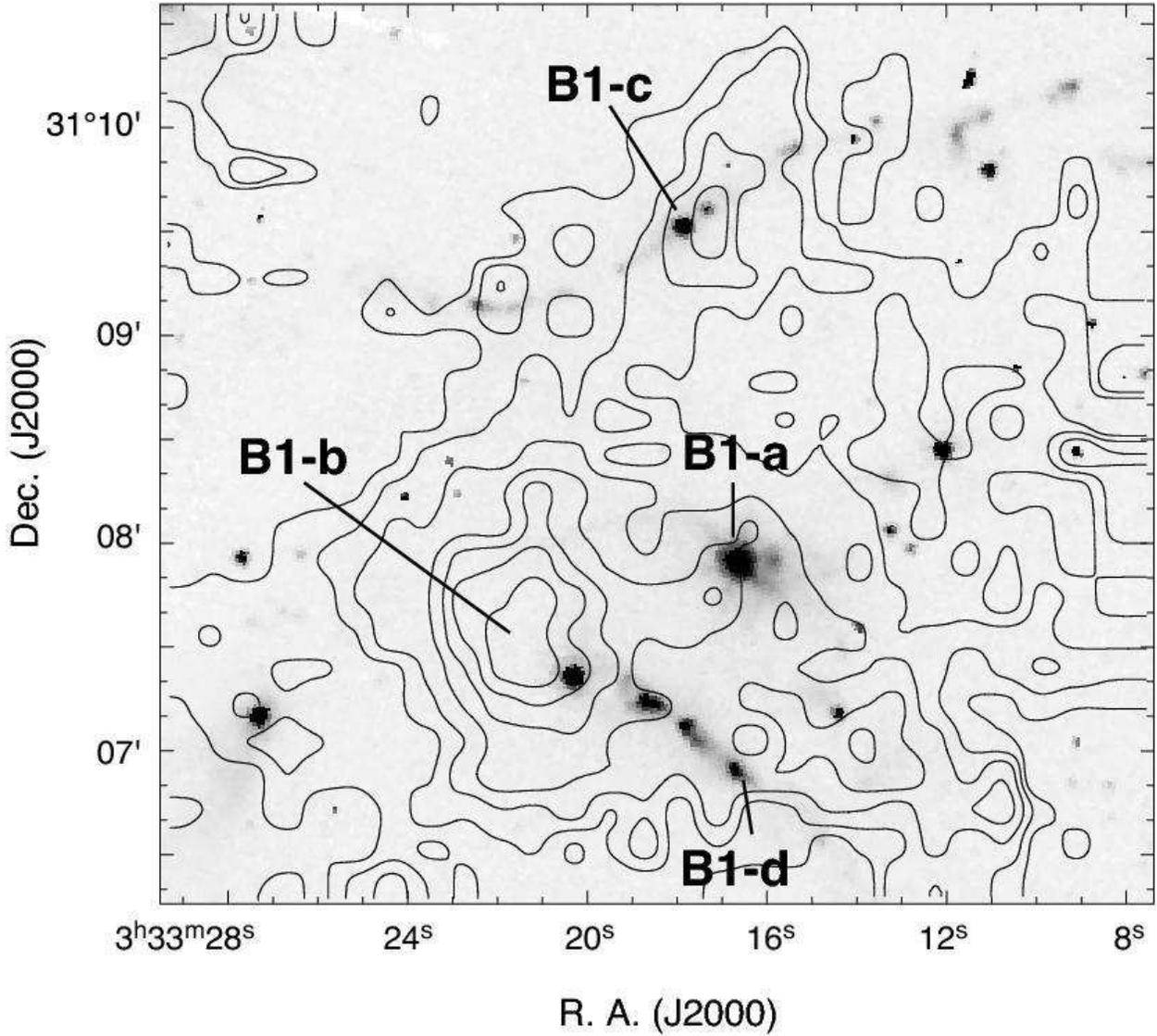}
\caption{Integrated H$^{13}$CO$^{+}$ (J=1-0) emission from B1 (contours) on top of the Spitzer IRAC band 2 (4.5 ${\mu}$m) image (grey scale). The velocity range of the H$^{13}$CO$^{+}$ map is from $V_{\rm LSR}$ = 5.0 to 8.0 km s$^{-1}$. The contours start from 0.67 K km s$^{-1}$ (3 $\sigma$) with an interval of 0.44 K km s$^{-1}$ (2 $\sigma$).  \label{fig1}}
\end{figure}

Figure \ref{fig1} shows an integrated intensity map of the H$^{13}$CO$^+$ (J=1-0) line emission from B1 region observed with the NRO 45m telescope. 
Since the H$^{13}$CO$^+$ J=1-0 line, having a critical density of 8$\times$10$^4$ cm$^{-3}$, is sensitive to high density gas, the H$^{13}$CO$^+$ map is frequently used to identify dense cores in molecular clouds \citep[e.g.][]{Oni02, Ike09}. 
The overall distribution of the H$^{13}$CO$^+$ emission is similar to those of NH$_3$ \citep {Bach90}  and dust continuum emission at 850 $\mu$m \citep{Matt02, Wala05, Hat05} and at 1.1 mm \citep{Eno06}.
The most prominent core in this region is B1-b, which has a size of ~100" x 60" (0.12 pc${\times}$0.07 pc) in FWHM. 
The northern peak corresponds to B1-c, which harbors a class 0 protostar with an S-shaped outflow seen in the {\it Spitzer} IRAC image.
This outflow is also seen in the near infrared \citep{Wala05}, and CO lines in millimeter \citep{Matt06, Hira10} and submillimeter \citep{Hat07}.
B1-a is located at ${\thicksim}$1' west of B1-b. 
B1-a harbors a low-luminosity (L ${\thicksim}$ 1.2 L$_{\sun}$\footnote{The luminosity of IRAS 03301+3057 was calculated to be $\sim$2.7 L$_{\sun}$ by assuming a distance of 350 pc (Hirano et al. 1997)). We have scaled this number to the value at a distance of 230 pc.} IRAS source, IRAS 03301+3057, which is driving an outflow dominated by the blueshifted component \citep{Hira97, Hira10}.
At the position of the submm source, B1-d, the H$^{13}$CO$^+$ map does not show significant enhancement.
Instead, the H$^{13}$CO$^+$ emission is enhanced at $\sim$30\arcsec west of B1-d.

Figure 2 shows the H$^{13}$CO$^+$ and HC$^{18}$O$^+$ spectra measured at the positions of B1-bN and B1-bS.
The HC$^{18}$O$^+$ emission was detected at the both positions.
The H$^{13}$CO$^+$/HC$^{18}$O$^+$  line ratios at these positions are $\sim$8 (see Table 3), which is close to the value of 5.5 expected for optically thin condition.
Therefore, the H$^{13}$CO$^+$ emission in B1-b, which is brightest in the mapped region, is likely to be optically thin.
Although the HC$^{18}$O$^+$ was not observed toward the other cores, we expect that the H$^{13}$CO$^+$ line is optically thin in the entire region of the mapped area.
The line profiles of H$^{13}$CO$^+$ and HC$^{18}$O$^+$ imply that there are two velocity components along the line of sight of B1-b; one is at ${\sim}$6.5 km s$^{-1}$ and the other is at $\sim$7.2 km s$^{-1}$.
At the position of B1-bS, the brightness temperatures of two velocity components are comparable, while at B1-bN, the 6.5 km s$^{-1}$ component is brighter than the 7.2 km s$^{-1}$ component.
The two velocity components in B1-b were also observed in the N$_2$H$^+$ and N$_2$D$^+$ $J$=3--2 by \citet{Huang13}.
In order to examine the spatial distributions of two velocity components, we made the maps of the H$^{13}$CO$^+$ in two velocity ranges (Figure \ref{fig3}).
The map of the lower velocity emission is similar to that of the total integrated intensity; this means that most of the H$^{13}$CO$^+$ emission from the B1 region is in this velocity range.
On the other hand, the emission in the higher velocity range mainly comes from the B1-b region.
This implies that there is another dense cloud having a radial velocity of $\sim$7.2 km s$^{-1}$ along the same line of sight of B1-b.
In addition to the two velocity components, the H$^{13}$CO$^+$ emission exhibits the redshifted wing seen in the velocity range of 8--10 km s$^{-1}$. 
This component is likely to be extended over the B1-b core, because  the wing is seen in the spectra of both B1-bN and B1-bS.
However, the spatial distribution of this component is unknown, because the mapping data do not have enough sensitivity to detect this component.

The column density of the H$^{13}$CO$^+$ molecule was estimated under the assumption of local thermodynamic equilibrium (LTE) and optically thin emission using the formula:
\[N({\rm H^{13}CO^+})= \frac{3k}{4{\pi}^3{\mu}^2{\nu}} \frac{kT_{\rm ex}}{h{\nu}} e^{\frac{h{\nu}}{kT_{\rm ex}}}\int{T_{\rm mb}}dv.\]
We assumed an excitation temperature $T_{\rm ex}$ of 12 K, which is same as the rotation temperature derived from the NH$_3$ observations \citep{Bach90}.
The LTE assumption is considered to be reasonable for the cores in B1, because the volume densities of the cores derived from the dust continuum observations are higher than 10$^5$ cm$^{-3}$ \citep [e.g.][]{Kirk07}.
The dipole moment of the H$^{13}$CO$^+$ molecule is adopted to be ${\mu}$ = 3.90 D from the CDMS catalog \citep{Mull05}.
The emission was integrated over the velocity range from $V_{\rm LSR}$ = 5.0 to 8.0 km s$^{-1}$.
The H$^{13}$CO$^+$ column densities at the positions of the B1-a, B1-b, and B1-c cores are calculated to be (2--4)$\times$10$^{12}$ cm$^{-2}$.
The LTE mass is defined as
\[M_{\rm core} = N({\rm H^{13}CO^+}) [X({\rm H^{13}CO^+})]^{-1} {\mu}_g m_{H_2} A.\]
where $A$ is the area within the half-intensity contour of the integrated intensity, ${\mu}_g$ is the mean atomic weight of 1.41.
We adopted $X(\rm {H^{13}CO^+}) = 8.3{\times}10^{-11}$ for the H$^{13}$CO$^+$ fractional abundance \citep{Fre87}.
The masses of the cores are estimated to be 2.0 $M_{\odot}$ for B1-a, 4.7 $M_{\odot}$ for B1-b, and 1.6 $M_{\odot}$ for B1-c.
Since the H$^{13}$CO$^+$ fractional abundance in the B1-b core is likely to be lower than the value adopted here (see Section 3.1.2), the mass derived here is considered to be the lower limit.

\subsection{Two continuum sources in B1-b}
Figure \ref{fig4} shows the continuum emission from B1-b region observed at 850 ${\mu}$m with the JCMT SCUBA and at 3 mm with the NMA overlaid on the Spitzer MIPS 24 ${\mu}$m image. 
It is shown that both B1-bN and B1-bS are detected at 3 mm with the NMA, but not in the Spitzer 24 $\mu$m image.
As shown in Figure \ref{fig5}, these two sources are clearly detected with the interferometers in all wavebands from 7 mm to 1.1 mm.
On the other hand, there is no continuum emission from the Spitzer source (hereafter, referred to as B1-bW\footnote{This Spitzer source is also called JJKM 39 by \citet{Wala09}}), although this source is detected in all 4 bands of IRAC and 24 and 70 $\mu$m bands of MIPS \citep{Jor06}.
The 7 mm map at 0\farcs5 resolution reveals that both two sources have spatially compact component (Figure \ref{fig6}).
The sizes of the compact components were derived from two dimensional Gaussian fit to the images.
The beam deconvolved size of B1-bN is 0\farcs42$\times$0\farcs23 (96$\times$54 AU) with a position angle of 24$^{\circ}$, and that of B1-bS is 0\farcs51$\times$0\farcs43 (116$\times$99 AU) with a position angle of 78$^{\circ}$.
The coordinates of the continuum peaks determined from the 7 mm map are R.A. (2000) = 3$^{\rm h}$ 33$^{\rm m}$ 21.2$^{\rm s}$, Dec. (2000) = 31$^{\circ}$ 07$'$ 43\farcs8 for B1-bN and $\alpha$(2000) = 3$^{\rm h}$ 33$^{\rm m}$ 21.4$^{\rm s}$, $\delta$(2000) = 31$^{\circ}$ 07' 26\farcs4 for B1-bS. 
There is no multiplicity in B1-bN nor B1-bS at a resolution of 0\farcs5.

The continuum fluxes of each source at different wavelengths are listed in Table 4.
For the {\it Spitzer} data at 24 and 70 ${\mu}$m, in which sources were not detected, 3 $\sigma$ levels were measured as the upper limits.
Table 4 includes the recent detection of far-IR emisison by {\it Herschel} \citep{Pez12}.
B1-bS was detected in the 70 ${\mu}$m PACS band, while B1-bN was not detected.
In the PACS 100 and 160 ${\mu}$m bands, two sources are clearly detected.
The 250, 350 and 500 ${\mu}$m data of SPIRE are not included in this table, because two sources are not separated well in these wavebands with lower angular resolutions.
The fluxes at 350 and 850 ${\mu}$m were measured using an aperture with a size of 20$''$$\times$20$''$.
The positions of two sources in the 350 ${\mu}$m map were shifted by $\sim$10$''$ southward from those in the maps of the other wavebands, probably because of the pointing problem.
Therefore, the flux values at 350 ${\mu}$m were measured by assuming that the locations of the two peaks at 350 ${\mu}$m are the same as those of the other wavebands.
The SMA, NMA, and VLA maps have been corrected for the primary beam response, and measured the fluxes using the two dimensional gaussian fitting task 'imfit'.
The submillimeter data at 350 and 850 $\mu$m were observed by single-dish telescopes, which can detect all the flux in the beam, including both compact and extended components. 
On the other hand, the longer wavelength data at 1.1, 1.3, 3 and 7 mm were observed by the interferometers, which are not sensitive to the extended structure. 
Since 1.1 mm continuum emission from the B1-b region has also been observed with Bolocam at the CSO \citep{Eno06}, we have compared the fluxes measured with Bolocam and the SMA.
The 1.1 mm map observed with the SMA was convolved to the same angular resolution of the Bolocam data, 31$''$.
With this angular resolution, single beam contains the flux from both B1-bN and B1-bS.
It is found that the SMA has filtered out approximately half of the total flux detected by Bolocam.

The spectral energy distributions (SEDs) of two sources are shown in Figure \ref{fig7}. 
The SEDs were fitted with the form of greybody radiation with a single emissivity, $\beta$, and a temperature, $T_{\rm dust}$,
\[F_\nu = \Omega B(\nu, T_{\rm dust})(1-e^{-{\tau}_{\nu}}) \] 
where F$_{\nu}$ is the flux density, $\Omega$ is the solid angle of the emitting region, B($\nu$, $T_{\rm dust}$) is the Plank function, $T_{\rm dust}$ is the dust temperature, and $\tau_\nu$ is the optical depth of dust that is assumed to be proportional to $\nu^\beta$. 
The derived parameters, $T_{\rm dust}$, $\beta$, $\tau_{\rm 230 GHz}$, and the source size $r = \sqrt{\ln {2}  \Omega / \pi}$ are listed in Table 5.
The SED of B1-bN was fitted by the dust temperature of $T_{\rm dust}{\sim}$16 K and ${\beta}{\sim}$  2.0, and that of B1-bS was fitted by $T_{\rm dust}{\sim}$18 Kand ${\beta}{\sim}$ 1.3. 
As compared to the $T_{\rm dust}$ and $\beta$ values derived for the entire B1-b core using the Herschel and SCUBA-2 data by \citet{Sad13}, $\sim$10.5 K, and $\sim$2, respectively, the temperatures of B1-bN and B1-bS are higher, and the $\beta$ value for B1-bS is lower.
The derived source sizes ($\sim$1\arcsec)  are larger than the source sizes measured in the 7 mm map, although the uncertainties are large.
The data points of B1-bN at 3.3 mm and 7 mm were not fitted well in Figure \ref{fig7}. 
The contribution of the free-free emission from shock-ionized gas at millimeter wave range is unclear, because there is no report of the centimeter emission.
If the free-free emission from B1-bN is comparable to that of the class 0 protostar L1448C(N), which is known to be an active outflow source, the expected flux densities at millimeter wave range is less than 1 mJy \citep{Hirano10}.
The possible contribution of the free-free emission is negligible at 3 mm, and is $\sim$50\% at 7 mm.
Therefore, the free-free emissions unlikely to be the source of the excess at 3 and 7 mm bands.
In addition, the data points at 1.1 and 1.3 mm for both sources are lower than the curves of the grey-body fit.
As mentioned above, the data points at millimeter wavebands were observed with the interferometers, which are more sensitive to the spatially compact components.
Therefore, the observed trend can be explained if the ${\beta}$ values in the compact component of these sources, especially B1-bN,  are smaller than those of the extended component.
The similar results with smaller ${\beta}$ in the compact component has also found in the very young protostar HH211 \citep{Lee07}.

\citet{Pez12} fitted the SEDs of the B1-b sources using two components; one is a compact blackbody component (i.e. ${\beta}$ = 0), and the other is an extended greybody component.
We also tried to reproduce the entire SEDs in the same manner as \citet{Pez12}, assuming the sizes of the compact components to be the same as those determined from the 7 mm data.
In the case of B1-bN, the temperature of the compact component has an upper limit of 23 K in order to meet the upper limits at 70 $\mu$m.
The fitting solutions are bimodal depending on the $\beta$ value of the extended component.
If ${\beta}$(ext)$>$ 1.82, the SED is fitted by $T_{\rm dust}$(ext)${\sim}$11 K and a size of $>$3.6\arcsec (with a large uncertainty of $>$8\arcsec).
The ${\beta}$(ext) smaller than 1.82 gives the solution with $T_{\rm dust}$(ext)${\sim}$14--15 K and a size of $\sim$1\arcsec$\pm$0.7\arcsec.
Since it is unlikely that the size of the extended component is only $\sim$1\arcsec, the low temperature solutions with ${\beta}$(ext)$>$ 1.82 could be more realistic.
The blue dashed curve in Figure \ref{fig7} is the example of the low-temperature solution for ${\beta}$(ext) = 2.0 and $T_{\rm dust}$(ext) = 10.6$\pm$1.4 K. 
The temperature of the compact component was set to 23 K.
The size of the extended component is $r$(ext) = 16\arcsec with a large uncertainty of 91\arcsec.
Although two component model can reproduce the SED at mm wave range, the size of the extended component cannot be constrained.
In the case of B1-bS, which has no bimodal problem, the parameters derived from 2 component model are $T_{\rm dust}$(ext)${\sim}$15--16 K, ${\beta}$(ext)$\sim$1.4--1.6, $r$(ext)$\sim$2.8\arcsec, and $T_{\rm dust}$(cmp)${\sim}$20--25 K.
In Figure \ref{fig7}, the curve for ${\beta}$(ext) = 1.5, which brings $T_{\rm dust}$(ext) = 15.8$\pm$6.8 K,  $r$(ext)$\sim$2.8$\pm$9.7, and $T_{\rm dust}$(cmp) = 24.2$\pm$7.8 K, is shown.
It is obvious that two component model does not fit the 7 mm data for B1-bS.
In this paper, we use the single-component greybody fitting results for simplicity.

The bolometric luminosity L$_{\rm bol}$ of each source was calculated by integrating the greybody curve.
The bolometric temperature was derived following the method of \citet{Mye93}.
The submillimeter-to-bolometric luminosity ratio (L$_{\rm sub}$/L$_{\rm bol}$), which is used to classify the object as a Class 0 source, were calculated with L$_{\rm sub}$ measured longward of 350 ${\mu}$m \citep{And93}. 
The mass of gas and dust associated with each source was estimated using the formula:
\[ M_{\rm env}=\frac{F_{\nu} D^2}{\kappa_{\nu}B_{\nu}(T_{\rm dust})},  \]
where $D$ is the source distance and $\kappa_{\nu}$ is the dust mass opacity, which is derived from $\kappa_{\nu}=\kappa_0 ({\nu}/{\nu_0})^{\beta}$ where ${\kappa_0}$ is calculated to be 0.01 cm$^2$ g$^{-1}$ at ${\nu_0}=$230 GHz \citep{Oss94}. 
The derived values. $L_{\rm bol}$, $T_{\rm bol}$, $L_{\rm smm}$/$L_{\rm bol}$, and  $M_{\rm env}$ for the two sources are listed in Table 6.
The two sources have extremely low bolometric temperatures of ${\sim}$ 20 K and large $L_{\rm submm}$/$L_{\rm bol}$ ratios greater than 0.1.
These values satisfy the class 0 criteria, which are $T_{\rm bol}$ ${\textless}$ 70 K and $L_{\rm smm}$/$L_{\rm bol}$ ${\textgreater}$ 0.005 \citep{And00}.
The continuum properties of B1-bN and B1-bS are close to the parameters of the archetypical Class 0 source VLA 1623 (15-20 K, {\textgreater}0.052) \citep{And93, Fro05}.

\subsection{Molecular line data observed with the interferometers}

\subsubsection{H$^{13}$CO$^{+}$ $J$=1--0}

The integrated intensity map of the H$^{13}$CO$^+$ observed with the NMA is shown in Figure 8a.
The H$^{13}$CO$^+$ emission is only seen around B1-bS, and not toward the other source  B1-bN.
In addition, there is no significant H$^{13}$CO$^+$ emission around B1-bW.
It is obvious that the NMA recovered only a small fraction of the H$^{13}$CO$^+$ line flux.
In the H$^{13}$CO$^+$ map with short spacing data (Figure 8b), the spatially extended emission is recovered.
However, there is no enhancement at the position of B1-bN.
The H$^{13}$CO$^+$ map observed with the NMA is significantly different from the N$_2$H$^+$ and N$_2$D$^+$ maps observed with the SMA, which clearly show two peaks at B1-bN and B1-bS \citep{Huang13}.
Since the H$^{13}$CO$^+$ line is optically thin in this region, the observed intensity distribution implies that the column density of H$^{13}$CO$^+$ is not enhanced at B1-bN.

The column density of H$^{13}$CO$^+$ was estimated using the combined H$^{13}$CO$^+$ map.
The integrated intensities of H$^{13}$CO$^+$ emission at B1-bN and B1-bS are 3.21 K km s$^{-1}$ and 5.74 K km s$^{-1}$, respectively.
The derived $N$(H$^{13}$CO$^+$) was 3.7$\times$10$^{12}$ cm$^{-2}$ at B1-bN and 6.6$\times$10$^{12}$ cm$^{-2}$ at B1-bS, under the assumptions of LTE, optically thin H$^{13}$CO$^+$ emission, and $T_{\rm ex}$ = 12 K.
The H$_2$ column density was estimated using the continuum data at 230 GHz.
The map was convolved to the same beam size as the H$^{13}$CO$^+$ map, 7\farcs7$\times$6\farcs5, and applied for the primary beam correction.
We used the dust temperature values listed in Table5 and the dust mass opacity of 0.01 cm$^2$ g$^{-1}$ at 230 GHz \citep{Oss94}.
The H$_2$ column density at B1-bN was estimated to be 2.0$\times$10$^{23}$ cm$^{-2}$, and that    at B1-bS was 2.8$\times$10$^{23}$ cm$^{-2}$.
The derived H$^{13}$CO$^+$ fractional abundance $X$(H$^{13}$CO$^+$) was $\sim$1.9$\times$10$^{-11}$ at B1-bN and $\sim$2.4$\times$10$^{-11}$ at B1-bS.
If we take into account that the SMA filtered out approximately half of the continuum flux, the fractional abundance of H$^{13}$CO$^+$ at the positions of two sources becomes $\sim$1$\times$10$^{-11}$, which is a factor of $\sim$8 lower than the canonical values of 8.3$\times$10$^{-11}$.
Such a low value could be explained if the H$^{13}$CO$^+$ line is subthermally excited.
However, it is unlikely for the cases of B1-bN and B1-bS; assuming the depths of these sources are comparable to the beam of the H$^{13}$CO$^+$ map, the volume density of these regions estimated from the H$_2$ column densities is (0.8--1.2)$\times$10$^7$ cm$^{-3}$, which is high enough to thermalize the H$^{13}$CO$^+$ $J$=1--0 line with a critical density of 8$\times$10$^4$ cm$^{-3}$.

\subsubsection{CO $J$=2--1}

The CO $J$=2--1 emission was detected  with the SMA in the velocity ranges from $V_{\rm LSR}$ = $-$1 to +13 km s$^{-1}$ and from 18 to 26 km s$^{-1}$.
Figure \ref{fig9} shows the velocity-channel maps of the CO $J$=2--1, and Figure \ref{fig10} shows the maps of the larger area with four velocity intervals; $V_{\rm LSR}$ = $-$1 -- 5 km s$^{-1}$ (blueshifted), 6--7 km s$^{-1}$ (cloud systemic velocity), 8--13 km s$^{-1}$ (redshifted), and 18--26 km s$^{-1}$ (high-velocity redshifted).

In the blueshifted velocity range, there are three compact emission components in the primary beam of the SMA; the peaks of these components are located at $\sim$6\arcsec southwest of B1-bN, $\sim$5\arcsec southeast of B1-bS, and $\sim$3\arcsec south of B1-bW.
The blueshifted emission near B1-bS appears in the wide velocity range from $V_{\rm LSR}$ = $-$1 to 4 km s$^{-1}$, while the blueshifted emission near B1-bN and B1-bW is observed in the narrow velocity range close to the cloud systemic velocity.
In addition, Fig. 10a exhibits bright CO emission components outside the primary beam.
The prominent feature seen at  $\sim$1\arcmin away from the field center is likely to be the blueshifted outflow from B1-a \citep{Hira97}.
In the cloud systemic velocity range ($V_{\rm LSR}$ = 6--7 km s$^{-1}$, Fig. 10b), the CO emission is elongated from northwest to southeast with a peak between B1-bN and B1-bS.
The CO emission in this velocity range is rather weak probably because the missing flux is significant.
In the redshifted velocity range (Figure 10c), the CO emission is spatially extended with a complicated structure.
There are two emission arcs extending to the north beyond the field of view of the SMA.
The eastern arc appears at the velocity channels of $V_{\rm LSR}$ = 8 and 9 km s$^{-1}$ and the western arc mainly at $V_{\rm LSR}$ = 9 km s$^{-1}$ channel.
The redshifted emission extending to the north is also seen in the single-dish CO $J$=3--2 map of \citet{Hat07}.
It is likely that two arcs seen in the SMA map correspond to the outer rims of the large scale redshifted lobe.
The velocity range of this extended emission, $V_{\rm LSR}$ = 8--10 km s$^{-1}$, corresponds to that of the H$^{13}$CO$^+$ wing emission.
Therefore, it is possible that the origin of the H$^{13}$CO$^+$ wings shown in Fig. 2 is related to the spatially extended CO component.
The CO emission also extends to the southeast of B1-bS and to the southwest of B1-bW, forming the X-shape.
The center of this X-shape does not coincide with neither of the sources in this region.
The CO emission disappears at $V_{\rm LSR}$ = 14 km s$^{-1}$ and reappears in the velocity range of $V_{\rm LSR}$ = 18--26 km s$^{-1}$.
In this high velocity range (Fig. 10d), all the emission is confined in a single knot at $\sim$6\arcsec north of B1-bW.

The spatial distribution of the blueshifted emission suggests that each source is associated with a compact outflow component.
The outflow activity in B1-bW  implies that this source is also a young stellar object (YSO), although the continuum emission  is not detected in the mm and submm wavebands.
The redshifted counterpart of each outflow is not clearly separated because of the complicated structure.
If we attribute the local intensity maxima of the redshifted component to the nearest sources, 
the brightest spot at  $\sim$7\arcsec northwest of B1-bS is likely to be the redshifted counterpart of the B1-bS outflow, and
the second brightest spot at $\sim$8\arcsec southwest of B1-bN might be the redshifted counterpart of the B1-bN outflow.
There is a small peak at $\sim$2\arcsec southeast  B1-bW, which is the same position as the peak of the blueshifted emission.
It is likely that B1-bS is driving a bipolar outflow with a blueshifted lobe to the southeast and the redshifted lobe to the northwest.
The channel map shows that the blueshifted component of this outflow has a velocity gradient with the higher velocity emission at the larger distance from the source.
The velocity gradient in the redshifted component is unclear because of the contamination of the spatially extended emission.
In the cases of the outflows from B1-bN and B1-bW, the bipolarity is not clear.

The origin of the extended arcs in the redshifted velocity range is not clear. 
The single-dish map of \citet{Hat07} implies that the extended redshifted lobe originates from B1-bS.
However, there is no blueshifted counterpart to the south of B1-bS.
The extended red lobe might be the monopolar outflow from B1-bS, which is similar to the extended monopolar outflow lobe in IRAS 16293-2422 \citep{Miz90}.
Another possibility is the outflow from B1-bN. 
B1-bN is located at the inner edge of the ring formed by two arcs.
If the axis of the outflow is close to the line of sight, the limb of the outflow cavity can be observed as a ring-like structure, as in the case of the B1-a outflow \citep{Hira97}.
In this case, the redshifted lobe is much more extended as compared to the blueshifted lobe.
It is also possible that the entire redshifted emission seen in the 8--13 km s$^{-1}$ originates from a single large outflow with an edge-on geometry driven by either B1-bN or B1-bS.
In this case, the two arcs in the north correspond to the cavity walls of the northern lobe, and the emission ridges extending to the southeast and southwest are the walls to the southern lobe.
The lack of the extended blueshifted emission could be explained by the spatial filtering of the interferometer.
Although the single-dish results do not show the spatially extended blueshifted CO emission toward B1-b \citep{Hat07,Hira10}, this scenario cannot be ruled out if the velocity of the blueshifted component is close to the cloud systemic velocity.
The last possibility is the outflow driven by the source outside of the primary beam of the SMA.
However, the single-dish  maps do not show clear evidence of blueshifted counterpart to the north of B1-b region (except the blue shifted lobe driven by B1-c \citep {Hat07,Hira10}).
In addition, there is no YSO candidate source to the north of B1-b region.

The origin of the high-velocity redshifted knot is also an issue.
As shown in Fig. 11a, the high-velocity redshifted knot is located along the line of the mid-IR jet-like feature extending to the southeast direction. 
In addition, the knot has its internal velocity gradient along the axis of the jet-like feature (Figures 11b).
These imply that the CO knot has the same origin as the jet-like feature.
This jet-like feature was also shown in the H${_2}$ image \citep{Wala09}, and was interpreted as a jet driven by B1-bW.
However, the high-velocity CO knot does not point to the location of B1-bW, implying that the CO knot and the jet-like feature do not originate from B1-bW.
{ B1-bN could be the candidate for the driving source, because the knot and jet also point to this source.
However, it is less likely because there is no counter jet nor blueshifted knot to the northeast of B1-bN.
}
Figure \ref{fig12} shows the P-V diagram along the major axis of the knot (P.A. = 53$^{\circ}$). 
This cut is also the axis of the jet-like feature.
The P-V diagram shows that the { velocity gradient in the knot is almost linear}.
If this linear velocity feature is the ^^ ^^ hubble flow'' that is often observed in the protostellar outflows such as HH211  \citep{Gue99,Lee07}, the driving source of this outflow is expected to be at the position where this linear velocity feature intersects with the line of the cloud systemic velocity.
The systemic velocity of the cloud at the southwest of B1-b is derived to be $V_{\rm LSR}{\sim}$6.5 km s$^{-1}$ from the H$^{13}$CO$^+$ spectra.
{ The velocity gradient in the knot was determined in three different methods; 1) the major axis of the emission component in the P-V diagram, 2) the linear fit of the intensity profiles at different velocities, and 3) the linear fit of the velocities derived from the line profiles.
Because the observed velocity feature slightly deviates from linear, the lines obtained with different methods did not agree with each other.
The magenta line in Figure \ref{fig12} was obtained from the first method.
The major axis was determined from two dimensional gaussian fitting of the P-V diagram.
This line intersects with the $V_{\rm sys}$ line at  $\alpha$(2000) $\sim$ 3$^h$33$^m$18.5$^s$, $\delta$(2000) $\sim$ 31$^\circ$07\arcmin13\farcs5, which is $\sim$2\arcsec west of the location of the H$_2$ knot labeled MH5 in \citet{Wala05,Wala09}.}
{ The velocity gradient derived from the second method also support the possibility of MH5.
The red dots in Figure  \ref{fig12} were obtained by calculating the intensity-weighted mean positions at different velocities.
The velocity gradient derived from this method is slightly steeper than the one obtained from the first method.
The results did not change if the if the positions were determined by fitting the intensity profiles with a single gaussian component.
This line intersects with the $V_{\rm sys}$ line at $\alpha$(2000) $\sim$ 3$^h$33$^m$19.0$^s$, $\delta$(2000) $\sim$ 31$^\circ$07\arcmin18\farcs5, which corresponds to $\sim$7\arcsec northeast of MH5.}
Although MH5 is not associated with the submm and mm continuum peak, it is located in the region of high column density \citep{Wala05}. 
Therefore, it is possible that the unknown YSO is embedded in this region.
However, no 24 $\mu$m source is seen in the {\it Spitzer} MIPS image.
{ On the other hand, the shallower velocity gradient was obtained from the third method.
The blue dots in Figure \ref{fig12} plots the intensity-weighted mean (i.e. the 1st moment) velocities.
These velocities are fitted by the straight line (blue dashed line in Figure \ref{fig12}), which is significantly different from the previous two lines.
The results did not change if the velocities were obtained by fitting the CO line profiles (in the velocity range from 15 to 30 km s$^{-1}$) with a single gaussian component.
The blue dashed line intersects with the $V_{\rm sys}$ line near the location of B1-d (JJKM36)}, which is the sub millimeter source located at the SW corner of Figure \ref{fig10} and Figure 11a.
The jet-like feature seen in the infrared is passing through this source.
In addition, another jet-like feature that is probably the counter jet is also seen in the SW of B1-d.
B1-d is known to have a CO outflow with the redshifted lobe to the northeast and the blueshifted lobe to the southwest \citep{Hat07}, which is consistent to the redshifted nature of the CO knot.
Therefore, B1-d is also a candidate for the driving source.
Since both MH5 and B1-d are located outside of the field of view of the SMA observations, detailed properties of these sources are unknown.

\subsubsection{$^{13}$CO and C$^{18}$O $J$=2--1}

The $^{13}$CO $J$=2--1 emission was detected in the velocity range of $V_{\rm LSR}$ = 5 -- 11 km s$^{-1}$.
The $^{13}$CO peaks at $\sim$6\arcsec northwest of B1-bS, and extends to the northeast, northwest, and southeast (Figure \ref{fig13}).
As shown in Figure \ref{fig14}, the $^{13}$CO emission is elongated along the NW-SE direction in the blueshifted and systemic velocity ranges, while it is along the NE-SW direction in the redshifted velocity range. 
The brightest $^{13}$CO peak appears in the redshifted velocity range.
This component is likely to be the counterpart of the red component of the B1-bS outflow.
The blueshifted peak at $\sim$7\arcsec southeast of B1-bN, the location of which coincides with that of the CO 2--1 peak, is considered to originate from the blue component of the B1-bN outflow.

The C$^{18}$O $J$=2--1 emission was detected in the velocity range from $V_{\rm LSR}$ = 5.75 to 8.25 km s$^{-1}$, which is same as the cloud systemic velocity.
However, The C$^{18}$O does not peak at neither B1-bN nor B1-bS, but between two sources.
The low-level emission extends along the NW to SE direction, which is same as the CO and $^{13}$CO in the systemic velocity.
The C$^{18}$O is also associated with B1-bW, although the emission does not peak at the source position.
The C$^{18}$O emission around B1-bW has a velocity of $V_{\rm LSR}$ = 6.1 km s$^{-1}$.

\subsubsection{Physical parameters of the CO outflows}

The physical parameters of the outflows were derived using the CO and $^{13}$CO data cubes corrected for the primary beam attenuations.
Under the LTE condition with an excitation temperature of 12 K \citep{Bach90}, the masses of the outflows were estimated from the equation,
\begin{displaymath}
M=1.48\times 10^{-6}  {\rm exp}(5.56/T_{\rm ex}) \frac{T_{\rm ex}+0.93}{{\rm exp}( -11.07/T_{\rm ex})} d_{kpc}^2\int S_\nu \frac{\tau}{1-e^{-\tau}} dv.
\end{displaymath}
We adopted an [H$_2$/CO] abundance ratio of 10$^4$ and a mean atomic weight of the gas of 1.41.
Since the $^{13}$CO emission was detected in the blue component of the B1-bN outflow, the red component of the B1-bS outflow, and the part of the extended redshifted component, the optical depth of the CO emission was obtained from the CO/$^{13}$CO line ratio.
The abundance ratio between CO and $^{13}$CO was assumed to be 77 \citep{Wil94}.
The CO emission was assumed to be optically thin for the other components.
The momentum, $P$, and the momentum rate, $F$ were estimated by
\begin{displaymath}
P=\sum m(v_i)v_i\mbox{  and  } F=P/t_d,
\end{displaymath}
where $m(v_i)$ is a mass at velocity $v_i$, and $t_d$ is a dynamical timescale.
The systemic velocities of B1-bN and B1-bS were determined to be $V_{\rm LSR}$ = 7.2 and 6.3 km s$^{-1}$, respectively, from the N$_2$D$^+$ observations \citep{Huang13}.
The systemic velocity for B1-bW was derived to be $V_{\rm LSR}$ = 6.1 km s$^{-1}$ from the C$^{18}$O $J$=2--1 spectrum at this position.
The dynamical time scale of the outflow, $t_d$, was calculated by $R/V_{\rm flow}$, where R is the length of the outflow lobe and $V_{\rm flow}$ is the outflow velocity.
Since the blueshifted emission comes from three spatially compact regions, we adopted the distances between the sources and the peak  positions as the lengths of the outflow lobes.
In the case of the redshifted component, we divided this into the three compact components and the extended component.
We assumed that the compact regions around the local intensity maxima correspond to the red lobes of the outflows from B1-bN , B1-bS, and B1-bW, and adopted the distances between the sources and the local intensity maxima as the lengths of the lobes.
The spatially extended component, which consists of two arcs extending to the north, was assumed to be originated from B1-bS and have a length of $\sim$40\arcsec (9200 AU).
The observed $V_{\rm max}$ is 7--8 km s$^{-1}$ for the compact components of the B1-bS outflow, and 3--5 km s$^{-1}$ for the B1-bN and B1-bW outflows.
However, $V_{\rm max}$ is affected by the sensitivity of the observations.
Therefore, the characteristic velocity calculated by $V_{\rm flow} = P/M$ was used for estimating the dynamical parameters.
The characteristic velocities of the outflows in this region are $V_{\rm flow}$ is $\sim$2--4 km s$^{-1}$ except for the high-velocity redshifted component.
The parameters of the outflows observed in the B1-b region estimated using these methods are listed in Table 7.
{ Since the high-velocity redshifted component has two candidates for its driving source, the lobe size, the dynamical time scale, and the momentum rate are calculated for both cases.}

The masses of the compact outflows around B1-bN, B1-bS, and B1-bW are in the orders of 10$^{-4}$--10$^{-3}$ $M_{\odot}$.
The momenta and momentum rates without inclination correction are in the orders of 10$^{-4}$--10$^{-3}$ $M_{\odot}$ km s$^{-1}$ and 10$^{-7}$--10$^{-6}$ $M_{\odot}$ km s$^{-1}$ yr$^{-1}$, respectively.
If we adopt the inclination angle of the outflow axis from the plane of the sky to be 32.7$^{\circ}$, which corresponds to the mean inclination angle from assuming randomly oriented outflows, the momenta and momentum rates increase by a factor of 1.9 and 1.6, respectively.

As discussed in the previous section, the distribution of the CO emission in this region is complicated, especially in the redshifted component.
In addition, the interferometric observations are always affected by the missing flux.
Therefore, the current assignment of the redshifted components to the driving sources is not conclusive.
The outflow parameters listed in Table 7 might change significantly if the different assignment is adopted.
On the other hand, the structure of the blueshifted component is rather simple; three compact components belong to the three sources in this region.
In the blueshifted component, the spatially extended emission is not significant.
Therefore, the uncertainties of the parameters should be much less for the blueshifted components, except the ones originate in the unknown inclination angles.

\section{DISCUSSION}

\subsection{Evolutionary stages of B1-bN and B1-bS}

The continuum data from 7 mm to 1.1 mm suggest that both B1-bN and B1-bS have already formed central compact objects.
The CO outflow activities observed in B1-bN and B1-bS also support the presence of central objects.
On the other hand, the SEDs of these sources are characterized by very cold temperatures of $T_{\rm dust} <$ 20 K, which is comparable to those of prestellar cores.
Futhermore, both sources are not detected in the 24 $\mu$m band of {\it Spitzer} MIPS, in which most of the known Class 0 protostars are detected.
These results imply that B1-bN and B1-bS are in the earlier evolutionary stage than most of the class 0 sources, and are probably candidates for the first core sources as proposed by \citet{Pez12}.
In the case of the first core, theoretical models predict the luminosity of $L$ ${\la}$ 0.1 $L_{\odot}$ \citep{Omu07,Sai08} or $L$ ${\la}$ 0.25 $L_{\odot}$ \citep{Comm12}.
On the other hand, the bolometric luminosities $L_{\rm bol}$ of B1-bN and B1-bS are 0.14 and 0.31 $L_{\odot}$, respectively.
Since the bolometric luminosities derived from the SEDs include the contribution of external heating by the interstellar radiation field, the internal luminosity of the source, $L_{\rm int}$, which is lower than the measured $L_{\rm bol}$ \citep[e.g.][]{Eva01} needs to be compared with the models.
The simplest way of estimating the internal luminosity is to use the empirical relation between $L_{\rm int}$ and 70 $\mu$m flux \citep{Dun08}.
Using this method, the $L_{\rm int}$ of B1-bS is estimated to be $\sim$0.03 $L_{\odot}$.
Since B1-bN is not detected at 70 $\mu$m, the upper limit of its internal luminosity is calculated to be 0.004 $L_{\odot}$.
However, these numbers of $L_{\rm int}$ might be underestimated; on the basis of the comparison between the 24 and 60 $\mu$m flux densities of Cha-MMS1 and those of the models of \citet{Comm12}, \citet{Tsi13} pointed out that the $L_{\rm int}$ derived from the empirical relation of \citet{Dun08} could underestimate the $L_{\rm int}$ by a factor of 3--7.
If this effect is taken into account, the $L_{\rm int}$ for B1-bS is estimated to be $\sim$0.1--0.2 $L_{\odot}$, and that for B1-bN to be $<$0.01--0.03 $L_{\odot}$.
The $L_{\rm int}$ of B1-bN falls into the range of the first core.
On the other hand, B1-bS is too luminous unless this source has an inclination angle (from the plane of the sky) smaller than 60$^{\circ}$.

In the case of the first core, an outflow with a wide opening angle and a low velocity of $\sim$5 km s$^{-1}$ is predicted from the Magneto-Hydrodynamic (MHD) model of \citet{Mac08}.
The velocity of the outflow from the first core is slow, because the Kepler velocity of the first core with a shallow gravitational potential is slow.
Once the protostar is formed, the newly formed protostar starts driving a well-collimated outflow.
The flow velocity from the protostar is faster, because the gravitational potential of the protostar is deeper.
In the case of the B1-bN outflow, the observed $V_{\rm max}$ is only 3--5 km s$^{-1}$, which is in the range of the outflow expected for the first core.
If the current assignment of the blue and red lobes to B1-bN is correct, the spatial overlap of the two lobes and the morphology of the blue lobe suggest that this outflow is viewed from nearly pole-on.
In such a case, the true velocity of the outflow is close to the observed radial velocity.
However, the current assignment of the redshifted components is not yet conclusive because of the complicated distribution of the redshifted emission.
On the other hand, the B1-bS outflow has a $V_{\rm max}$  of 7--8 km s$^{-1}$ without inclination correction, which is a little bit larger than that of the first core predicted by \citet{Mac08}.
In addition, the blue and red lobes are spatially separated in this outflow, suggesting that the outflow axis is inclined; if the inclination angle from the plane of the sky is 32.7$^{\circ}$ (the mean inclination angle of the randomly oriented outflows), the true velocity is a factor of 2 larger than the radial velocity. 
To summarize, the slow velocity of 3--5 km s$^{-1}$ in B1-bN outflow is consistent with the property of the first core, while the larger velocity of the B1-bS outflow implies the presence of the central object more massive than the first core.

The outflow properties of the very-low luminosity ($\lesssim$0.1 $L_{\odot}$) sources including four first core candidates and three very-low luminosity protostars have been summarized by \citet{Dun11}.
The found that both the first core candidates and the very-low luminosity protostars are driving the low velocity ($\leq$3.5 km s$^{-1}$) outflows (except for the first core candidate source L1448 IRS2E), and that the physical parameters of these outflows such as mass, momentum, and momentum rate, show significant dispersion of several orders of magnitude.
There is no obvious correlation between the outflow parameters and the source luminosity probably because of the small number of the sample.
In addition, there is no difference between the outflows from the first core candidates and those from the very-low luminosity protostars.
The characteristic velocities of the B1-bN and B1-bS outflows, 2--4 km s$^{-1}$, are comparable to those of the first core candidates as well as the very-low luminosity protostars.
The masses, momenta, and momentum rates of the B1-bN and B1-bS outflows are also in the range of those of the outflows from first core candidates and very-low luminosity protostars.

As discussed above, B1-bS has slightly larger internal luminosity and outflow velocity as compared to those of the first core, suggesting that this source is in the more evolved stage.
However, B1-bS is much less luminous as compared to the theoretically predicted luminosity of the second core source, which is expected to jump up $\gtrsim$1 $L_{\odot}$ immediately after the second collapse \citep{Sai08}.
If B1-bS has already formed a second core, i.e. protostar, the lower luminosity of this source is probably because the mass accretion onto the central protostar of this source is currently inactive.

The 7 mm data revealed that the size of the compact objects in these sources is $\sim$100 AU, which is comparable to the size of the compact objects observed in the Class 0 sources such as L1448C(N) \citep{Hirano10} and HH212 \citep{Lee08}.
The compact structures observed in L1448C(N) and HH212 are considered to be the circumstellar disks because they are elongated perpendicular to the outflow axes.
In the case of B1-bS, the position angle of the CO outflow is along the NW-SE direction, while the 7 mm continuum image is elongated along the NE-SW direction, which is roughly perpendicular to the outflow axis.
Therefore, the compact component in B1-bS is considered to be the origin of the circumstellar disk.
In the case of B1-bN, the relation between the source elongation and CO outflow is unclear, because the CO outflow does not show clear bipolarity.
However, it is unlikely that the properties of the compact components in B1-bN and  B1-bS are significantly different. 

\subsection{Depletion of H$^{13}$CO$^+$ and its implication to the evolutionary stages of two sources}

The H$^{13}$CO$^+$ $J$=1--0 line is often used as a tracer of dense gas surrounding protostars.
In the case of Class 0 protostar such as B335 and L1527 IRS, the H$^{13}$CO$^+$ emission peaks at the positions of protostars and show elongation perpendicular to the outflow axes \citep{Sai99, Sai01}.
On the other hand, the H$^{13}$CO$^+$ does not peak at B1-bN; rather, it tends to avoid the position of B1-bN.
Such a disagreement in spatial distribution between the H$^{13}$CO$^+$ and dust continuum emission is also observed in the prestellar core, L1544 \citep{Cas02}.
Although the H$^{13}$CO$^+$ peaks at the position of B1-bS, it does not show centrally peaked distribution as compared to the dust continuum emission.
As described in section 3.3.1, this is because the fractional abundances of H$^{13}$CO$^+$ toward B1-bN and B1-bS are significantly lower than that of the canonical value.

Since the primary formation route of HCO$^+$ is H$_3^+$ + CO $\rightarrow$ HCO$^+$ + H$_2$, the low fractional abundance of H$^{13}$CO$^+$ implies that the significant fraction of CO is depleted onto dust grains in dense gas surrounding two sources \citep[e.g.][]{Jor04,Ber07}.
The CO depletion is also supported by the high N$_2$D$^+$/N$_2$H$^+$ ratios of $\sim$0.2 at the positions of B1-bN and B1-bS \citep{Huang13}.
The depletion of carbon-bearing molecules into dust grains increases the H$_2$D$^+$/H$_3^+$ ratio in the gas phase, resulting in the increase of deuteriated species such as N$_2$D$^+$ \citep{Roberts00}.
In addition, the spatial distribution of C$^{18}$O, which does not peak at B1-bN nor B1-bS,  also support that CO is deficient in the dense gas around two sources.
Although the C$^{18}$O emission is detected in the region between two sources, the spatial distribution of the C$^{18}$O in B1-b is significantly different from that of the class 0 protostar B335, in which the C$^{18}$O clearly peaks at the position of the protostar \citep{Yen11}.

The H$^{13}$CO$^+$ peak at B1-bS suggests that some fraction of CO has already returned to the gas phase probably because of the heat from the central source or the dynamical interaction between the outflow and ambient gas.
On the other hand, the gas around B1-bN is likely to be colder than the CO sublimation temperature of 20K.
These chemical properties support the conclusion derived from the mid to far-infrared SEDs, i.e. B1-bN is in the earlier evolutionary stage than B1-bS.

\subsection{Properties of B1-bW}

B1-bW is the brightest source in B1-b region in the {\it Spitzer} IRAC and MIPS images.
Because the MIR colors of this source is  consistent with an embedded YSO \citep{Jor06},  and because the CO outflow localized to this source is observed, it is likely that B1-bW is also a YSO embedded in the B1-b core.
The C$^{18}$O emission is also associated with this source.
On the other hand, this source is not detected in the mm and submm continuum emission.
The 3 $\sigma$ upper limit of the continuum flux at $\nu_0$ = 230 GHz was $\sim$7 mJy beam$^{-1}$.
The upper limit of the mass is estimated to be 0.01 $M_{\odot}$ for a dust temperature of 15 K and a dust mass opacity of 0.01 cm$^2$ g$^{-1}$ at ${\nu}_0$ = 230 GHz \citep{Oss94}.
Although the dust temperature of this source is unknown, the higher dust temperature lowers the upper limit.
The mass of the circumstellar material surrounding B1-bW is more than 30--40 times less than those of B1-bN and B1-bS.
Due to the small amount of circumstellar material, the central star of B1-bW is likely to be less obscured in the near and mid infrared wavebands.
This is similar to the case of L1448C(S), which is surrounded by small amount of circumstellar material of less than 0.01 $M_{\odot}$ and is also bright in mid infrared \citep{Hirano10}.
In the case of L1448C(S), the small amount of circumstellar material is probably because the envelope gas has been stripped off by the powerful outflow from nearby class 0 source L1448C(N) \citep{Hirano10}.
Similar scenario is also possible for the case of B1-bW.
If the high-velocity CO knot is part of the jet/outflow from either MH5 or B1-d, this jet/outflow could plunge into dense gas in the B1-b core. 
The caved structure open to the west seen in the redshifted CO map (Figure \ref{fig11}) also support such an interaction.
Since B1-bW is located at the southwestern edge of the B1-b core, it is possible that the outflowing gas has stripped away the dense gas surrounding B1-bW.

\section{CONCLUSIONS}

Our main conclusions are summarized as follows.
\begin{enumerate}
\item The H$^{13}$CO$^+$ $J$=1--0 map observed with the NRO 45m  telescope has revealed that B1-b is the most prominent core in the B1 region.
The size and mass of the B1-b core were estimated to be 0.12 pc $\times$ 0.07 pc in FWHM and 4.7 $M_{\odot}$, respectively.
\item  The B1-b core contains two submm/mm continuum sources, B1-bN and B1-bS.
The SEDs of these sources are characterized by very cold temperatures of $T_{\rm dust}<$20 K.
The bolometric luminosity of these sources are 0.14--0.31 $L_{\odot}$.
If the contribution of external heating by the interstellar radiation field is taken into account, the internal luminosity $L_{\rm int}$ of B1-bN is estimated to be $<$ 0.01--0.03 $L_{\odot}$, and that of B1-bS to be $\sim$0.1--0.2 $L_{\odot}$.
\item Both B1-bN and B1-bS are associated with the CO outflows with low characteristic velocities of 2--4 km s$^{-1}$.
The maximum velocities of the B1-bN and B1-bS outflows are 3--5 km s$^{-1}$ and 7--8 km s$^{-1}$, respectively.
\item The SEDs and the outflow properties suggest that B1-bN and B1-bS are in the earlier evolutionary stage than most of the known class 0 protostars.
Especially, the internal luminosity and outflow velocity of B1-bN agree with those of the first core predicted by the MHD simulation \citep{Comm12,Mac08,Sai08}.
On the other hand, B1-bS has slightly larger internal luminosity and outflow velocity as compared to those of the first core, suggesting that this source could be more evolved than the first core.
\item The continuum data from 7 mm to 1.1 mm observed with the interferometers suggest that both B1-bN and B1-bS have already formed central compact objects.
The size of the compact objects in these sources is$\sim$100 AU, which is comparable to that of the compact objects in Class 0 sources.
These compact objects might be the origin of the circumstellar disks.
\item The H$^{13}$CO$^+ $line does not show significant enhancement at the positions of B1-bN. 
Although the H$^{13}$CO$^+$ peaks at the position of B1-bS, it does not show centrally peaked distribution.
The fractional abundance of H$^{13}$CO$^+$ was estimated to be $\sim$(1--2)$\times$10$^{-11}$ at both B1-bN and B1-bS, which is a factor of 4--8 lower than the canonical value.
\item The low fractional abundance of H$^{13}$CO$^+$ implies that significant fraction of CO is depleted onto dust grains in dense gas surrounding B1-bN and B1-bS. The CO depletion is also supported by the high D/H ratio derived from the observations of N$_2$H$^+$ and N$_2$D$^+$ \citep{Huang13}.
The chemical properties also support that B1-bN and B1-bS are in the early stage of protostellar evolution.
\item The CO emission seen in the high-velocity range ($\sim$12--20 km s$^{-1}$ redshifted with respect to the cloud systemic velocity) is confined in a single knot along the line of the mid-IR jet-like structure. 
The knot and jet are likely to be originate from  the protostar located to the southwest of B1-b.
The possible candidate is { an unknown source in the H$_2$ knot MH5 or the other submillimeter source B1-d.}
\item The {\it Spitzer} source, B1-bW, also shows the outflow activity. 
It is likely that B1-bW is also a YSO embedded in the B1-b core.
The non detection of B1-bW in the mm and submm continuum suggest that the circumstellar material surrounding this source is less than 0.01 $M_{\odot}$.
The small amount of circumstellar material abound B1-W is probably because the outflow from the southwest, observed as the mid-IR jet-like structure, may have stripped away the dense gas surrounding B1-W.
\end{enumerate}

\acknowledgments

We wish to thank the staff of SMA and NRO for the operation of our observations and for support in data reduction. 
We thank to Dr. S.-P. Lai for helping in data reduction of the {\it Spitzer} data.
N. H. thanks  to Dr. T. Umemoto for his help in the observations of the 45m telescope.
N. H. is supported by NSC grant 102-2119-M-001-009-MY2.

\clearpage

\begin{deluxetable}{cccccc}
\tabletypesize{\small}
\tablecolumns{5}
\tablewidth{0pc}
\tablecaption{Interferometric observations -- continuum}
\tablehead{\colhead{$\lambda$} & \colhead{$\nu$} & \colhead{Telescope} & \colhead {beamsize}& \colhead{PA} & \colhead{rms}\\
\colhead{(mm)} & \colhead{(GHz)} & \colhead{} & \colhead{} & \colhead{} & \colhead{(mJy beam$^{-1}$)}}
\startdata
7.0 & 43.31, 43.36 & VLA & 0\farcs51$\times$0\farcs45 & $-$71$^{\circ}$ & 0.11 \\
3.3 & 86.75, 98.75 & NMA & 3\farcs3$\times$2\farcs9 & $-$73$^{\circ}$ & 1.4 \\
1.3 & 220.4, 230.4 & SMA & 6\farcs3$\times$4\farcs0 & 49$^{\circ}$ & 2.3 \\
1.1 & 278.8, 288.8 & SMA & 3\farcs3$\times$2\farcs9 & 57$^{\circ}$ & 3.1 \\
\enddata
\label{table1}
\end{deluxetable}

\begin{deluxetable}{lcccccc}
\renewcommand{\arraystretch}{0.85} 
\tabletypesize{\small}
\tablecolumns{7}
\tablewidth{0pc}
\tablecaption{Interferometric observations -- lines}
\tablehead{\colhead{Line} & \colhead{Frequency} & \colhead{Telescope} & \colhead {beamsize}& \colhead{PA} & \colhead{rms} & \colhead{$\Delta$V} \\
\colhead{} &  \colhead{(GHz)} & \colhead{} & \colhead{} & \colhead{} & \colhead{(Jy beam$^{-1}$)} & \colhead{(km s$^{-1}$)}}
\startdata
H$^{13}$CO$^+$ 1--0 & 86.7542884 & NMA  & 7\farcs2$\times$6\farcs1 & $-$23$^{\circ}$ & 0.15 & 0.25 \\
 & & NMA+45m & 7\farcs7$\times$6\farcs5 & $-$23$^{\circ}$ & 0.14 & 0.25 \\
C$^{18}$O  2--1 & 219.5603541 & SMA & 6\farcs5$\times$4\farcs3 & 36$^{\circ}$ & 0.12 & 0.25 \\
$^{13}$CO  2--1 & 220.3986842 & SMA & 6\farcs5$\times$4\farcs3 & 36$^{\circ}$ & 0.05 & 1.0 \\
CO  2--1  & 230.53797 & SMA  & 6\farcs2$\times$3\farcs9 & 49$^{\circ}$ & 0.1 & 1.0 \\
\enddata
\label{table2}
\end{deluxetable}

\clearpage

\clearpage

\begin{figure}
\epsscale{0.8}
\plotone{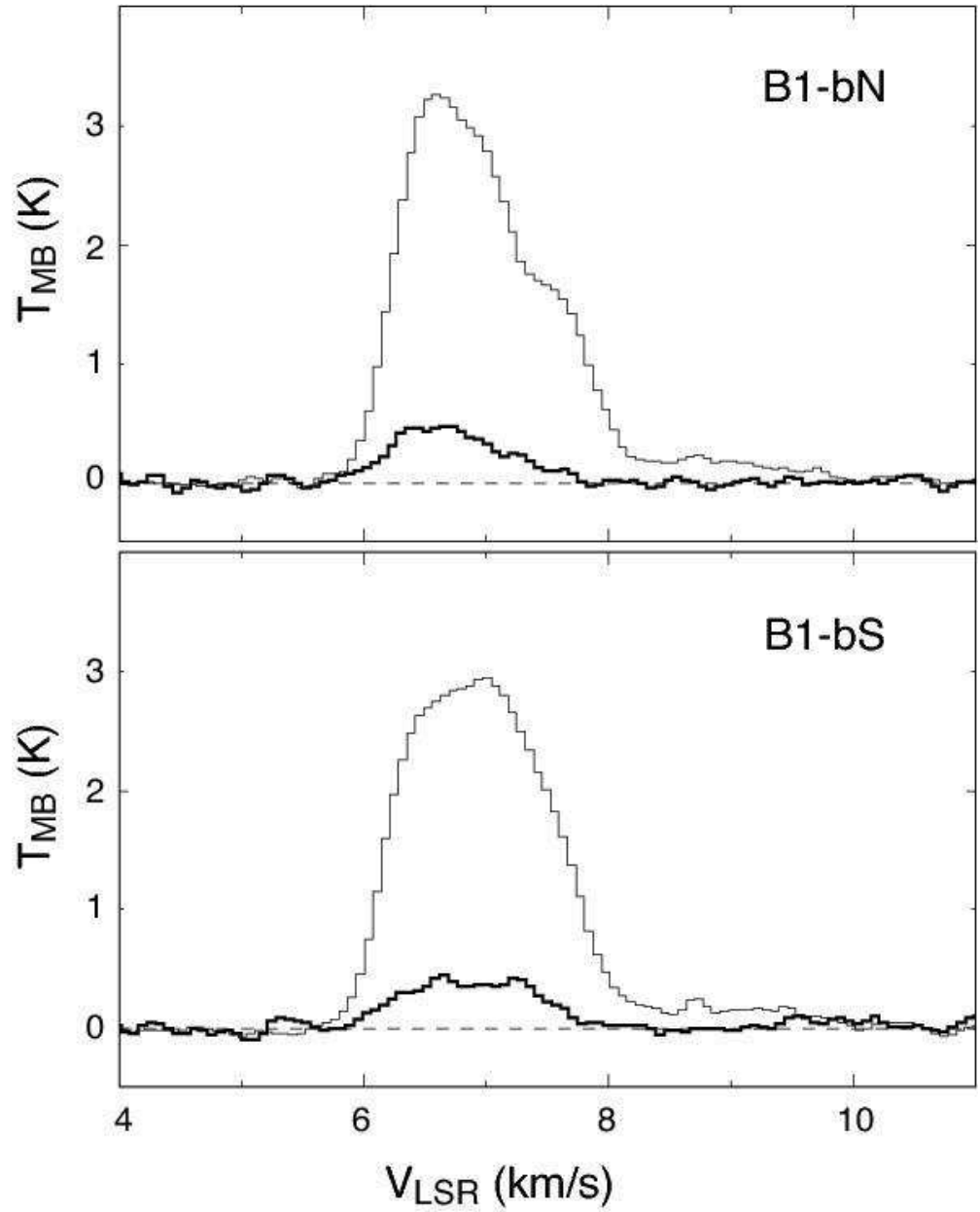}
\caption{H$^{13}$CO$^+$ $J$=1--0 (thin lines) and HC$^{18}$O$^+$ $J$=1--0 (thick lines) spectra toward the positions of B1-bN (top) and B1-bS (bottom).
\label{fig2}}
\end{figure}

\clearpage 

\begin{deluxetable}{lccccc}
\tabletypesize{\small}
\tablecolumns{6}
\tablewidth{0pc}
\tablecaption{Parameters of H$^{13}$CO$^+$ and HC$^{18}$O$^+$ at B1-bN and B1-bS}
\tablehead{\colhead{} &\colhead{} &\colhead{} & \colhead{${\int}T_{\rm mb}$(H$^{13}$CO$^+$)$dv$} &\colhead{${\int}T_{\rm mb}$(HC$^{18}$O$^+$)$dv$} &\colhead{H$^{13}$CO$^+$/HC$^{18}$O$^+$} \\
\colhead{} &\colhead{${\alpha}$(2000)} &\colhead{${\delta}$(2000)} &\colhead{K km s$^{-1}$} &\colhead{K km s$^{-1}$} &\colhead{}}

\startdata
B1-bN & 3$^{\rm h}$33$^{\rm m}$21.2$^{\rm s}$ & 31$^{\circ}$07$'$43\farcs8 & 4.33$\pm$0.02 & 0.51$\pm$0.02 & 8.5 \\
B1-bS & 3$^{\rm h}$33$^{\rm m}$21.4$^{\rm s}$ & 31$^{\circ}$07$'$26\farcs4 & 4.36$\pm$0.02 &0.56$\pm$0.02	& 7.9 \\
\enddata
\label{table 2}
\end{deluxetable}

\clearpage      

\begin{figure}
\epsscale{0.8}
\plotone{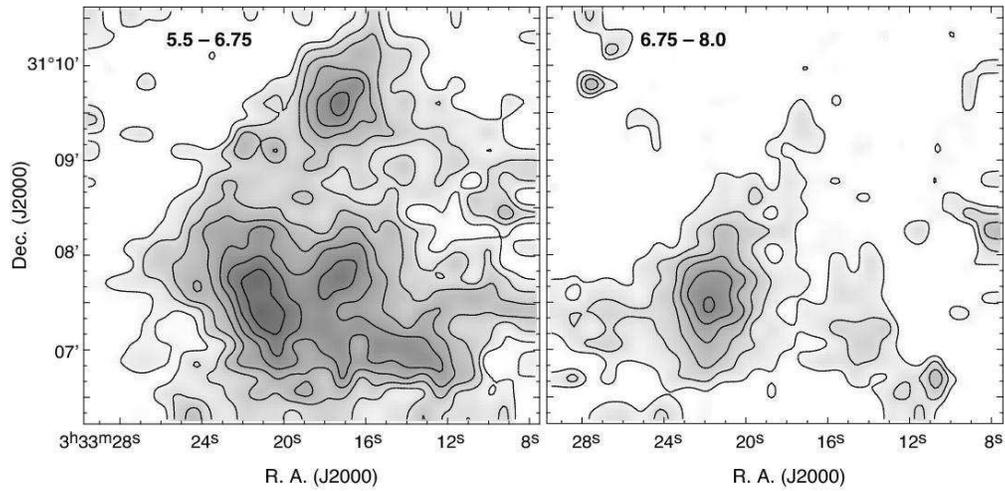}
\caption{H$^{13}$CO$^+$ $J$=1--0 in two velocity ranges ({\it Left}: $V_{\rm LSR}$ = 5.5--6.75 km s$^{-1}$, {\it Right}: $V_{\rm LSR}$ = 6.75--8.0 km s$^{-1}$). The contours start from 0.3 K km s$^{-1}$ (3 $\sigma$) with an interval of 0.2 K km s$^{-1}$ (2 $\sigma$).
\label{fig3}}
\end{figure}

\clearpage               

\begin{figure}
\plotone{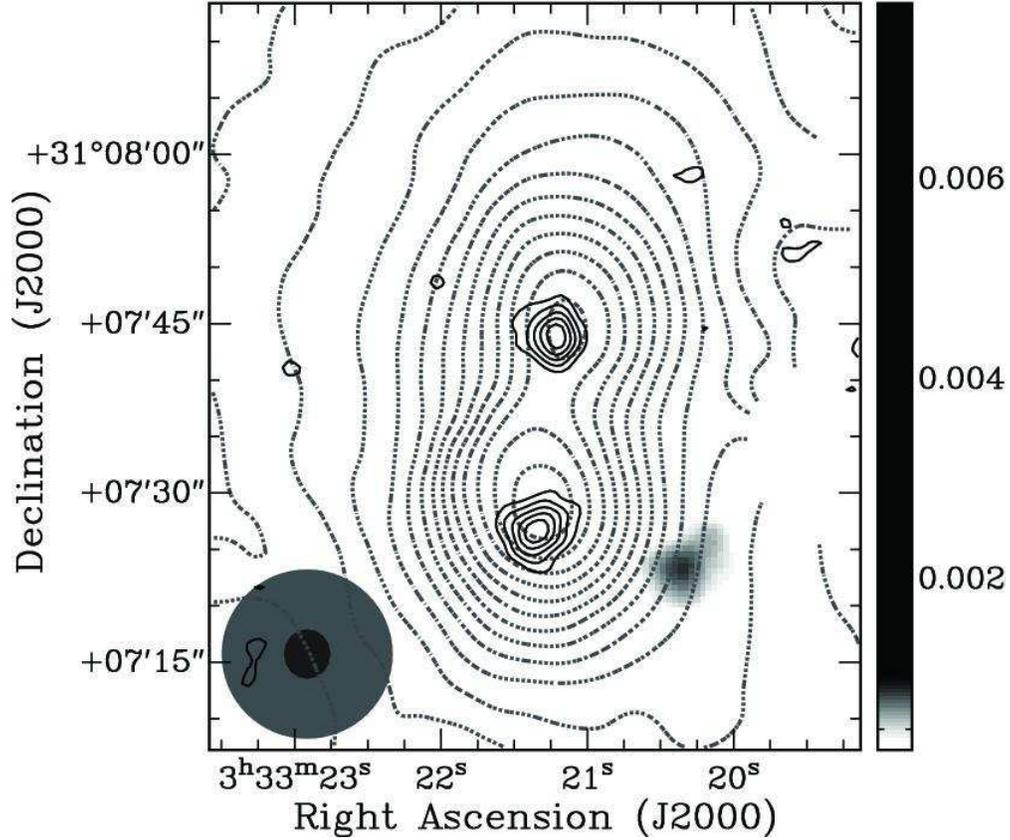}
\caption{850 ${\mu}$m (dot contours) and 3.3 mm (solid contours) continuum maps of the B1-b region on top of the Spitzer MIPS 24 ${\mu}$m image (grey scale). The 3.3 mm continuum has been applied for the primary beam correction. The gray-filled and black-filled circles in the lower-left courner are the beams of 850 ${\mu}$m and 3.3 mm, respectively. The contours are drawn every 0.08 Jy/beam (3 $\sigma$) for 850 ${\mu}$m and 3.4 mJy/beam (2.5 $\sigma$) for 3.3 mm. 
\label{fig4}}
\end{figure}

\clearpage 

\begin{figure}
\epsscale{1}
\plotone{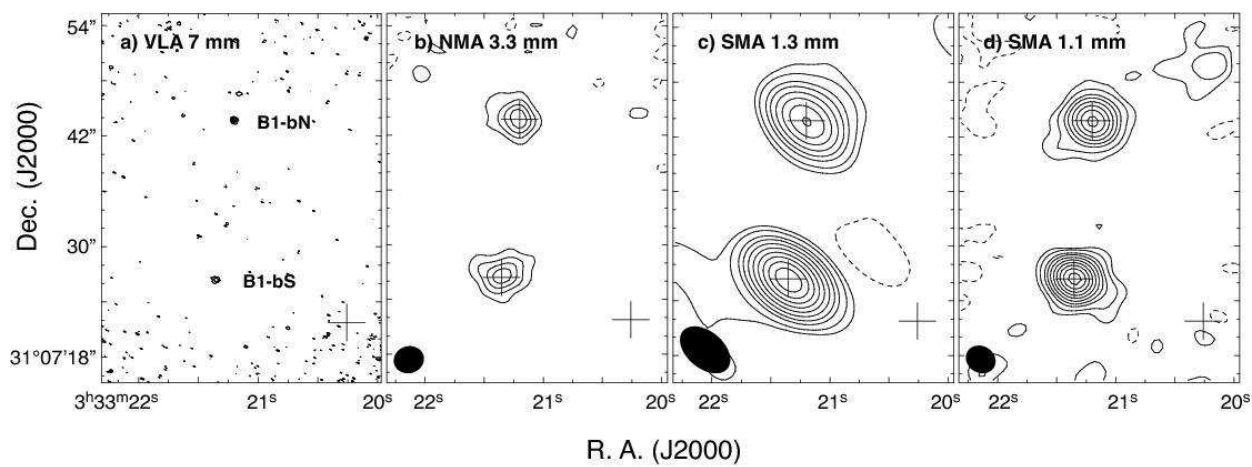}
\caption{Maps of the continuum emission at millimeter wavebands. All the maps are corrected for the primary beam response. (a) 7 mm continuum map observed with the VLA. The contours are drawn every 3 $\sigma$. The cross denotes the position of B1-bW. (b) 3.3 mm continuum map observed with the NMA. The contours are drawn every 3 $\sigma$. The crosses indicate the positions of B1-bN, B1-bS, and B1-bW. (c) 1.3 mm continuum map observed with the SMA. The contours start at 3$\sigma$ and are drawn at 6, 12, 18, 27, 39, 54, 72, 93, 117, and 144$\sigma$. (d) 1.1 mm continuum map observed with the SMA. The contours start at 3$\sigma$ and are drawn at 6, 12, 18, 27, 39, 54, 72, 93,  and 117$\sigma$. 
\label{fig5}}
\end{figure}

\clearpage

\begin{figure}
\plotone{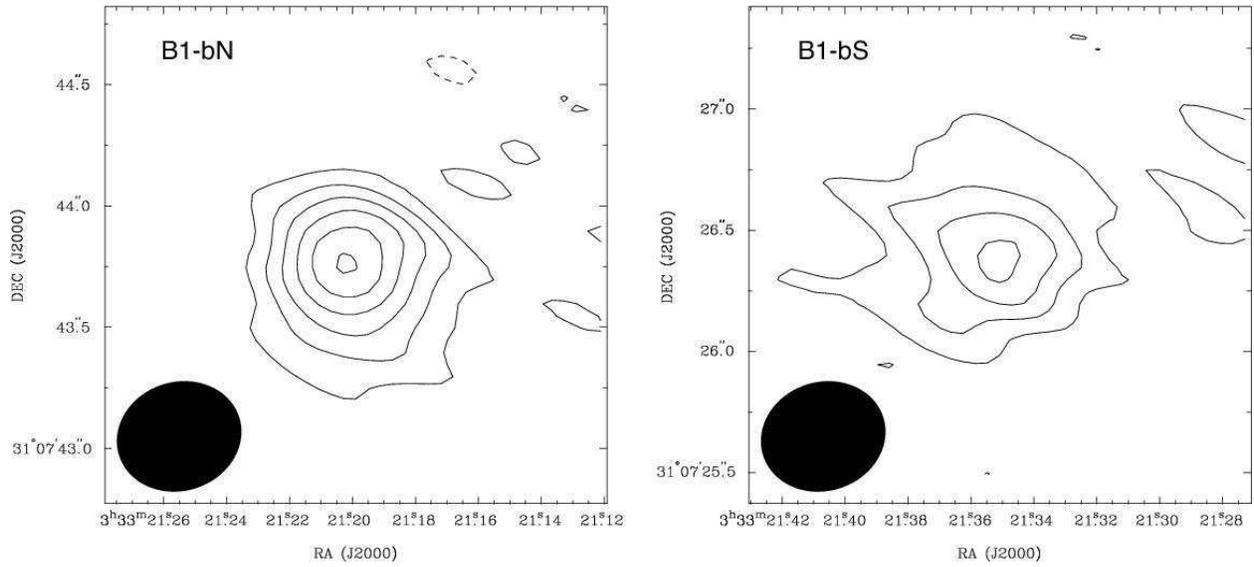}
\caption{7 mm continuum images of B1-bN ({\it left}) and B1-bS({\it right}). Contours are drawn every 2$\sigma$ intervals with the lowest contours at 2$\sigma$.
\label{fig6}}
\end{figure}

\clearpage 

\begin{deluxetable}{ccccc}
\small
\renewcommand{\arraystretch}{1} 
\tablecolumns{5}
\tablewidth{-10pc}
\tablecaption{Photometric Data of B1-bN and B1-bS}
\tablehead{ 
\colhead{Wavelength} & \colhead{Flux Density (bN)}   & \colhead{Flux Density (bS)}   & \colhead{Error}  & \colhead{Aperture} \\
\colhead{(\micron)} & \colhead{(Jy)}     & \colhead{(Jy)} & \colhead{(Jy)} & \colhead{(arcsec)}}
\startdata
24 & \textless 2.2$\times$10$^{-4}$  & \textless 6.4$\times$10$^{-3}$  & \nodata  & 7 \\
70 & \textless 2.0$\times$10$^{-2}$  & \textless 9.4$\times$10$^{-2}$ & \nodata  & 16\\
70$^*$ &  \textless 3.7$\times$10$^{-2}$  & 1.7$\times$10$^{-1}$ & \nodata/0.11 & 6.9/4.4\\
100$^*$ & 5.8$\times$10$^{-1}$ & 2.24 & 0.29/0.25 & 8.5/3.4\\
160$^*$ & 2.95 & 9.1 & 0.73/1.2 & 8.9/5.8\\
350 & 6.0  & 7.2 & 0.3 & 20\\
850 & 1.03  & 1.24 & 0.03 & 20\\
1057 & 3.14$\times$10$^{-1}$ & 4.68$\times$10$^{-1}$ & 1.5$\times$10$^{-2}$ & \nodata$^{**}$\\
1300 & 1.92$\times$10$^{-1}$& 3.45$\times$10$^{-1}$ & 1.0$\times$10$^{-2}$ &  \nodata$^{**}$\\
3300 & 2.59$\times$10$^{-2}$  & 3.07$\times$10$^{-2}$ & 3.5$\times$10$^{-3}$  &  \nodata$^{**}$\\
7000 & 1.94$\times$10$^{-3}$ & 1.69$\times$10$^{-3}$ & 2.8$\times$10$^{-4}$ &  \nodata$^{**}$\\
\enddata
\tablenotetext{*}{Herschel PACS data from \citet{Pez12} after the color correction}
\tablenotetext{**}{Estimated from two-dimensional Gaussian fittings to the image} 
\label{table 3}
\end{deluxetable}
\clearpage

\begin{figure}
\epsscale{0.7}
\plotone{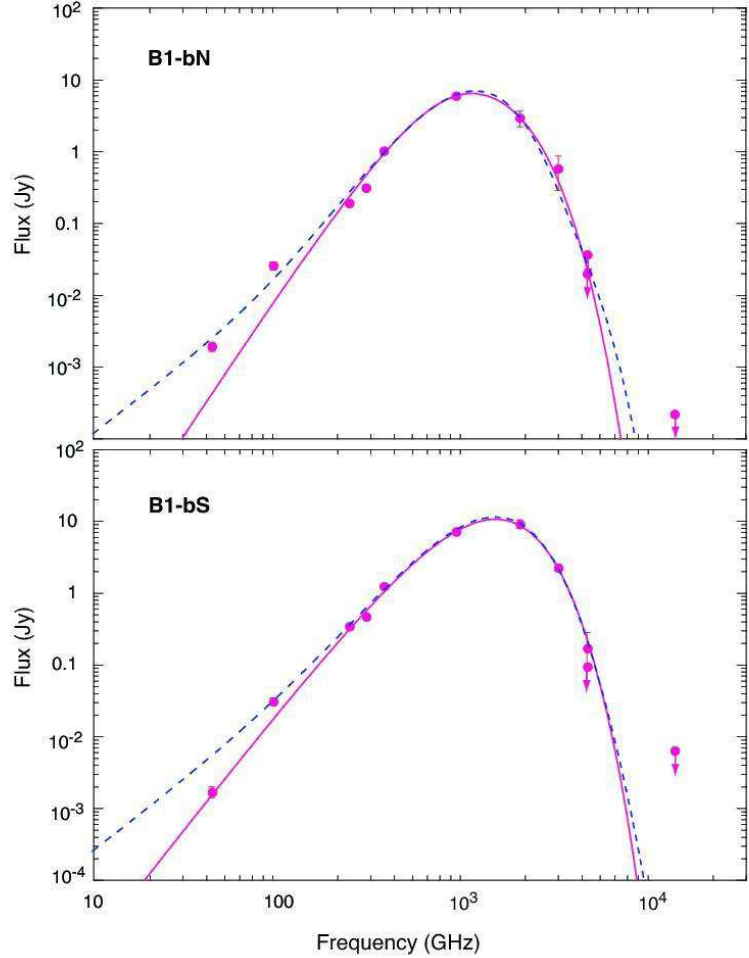} 
\caption{Spectral energy distributions of B1-bN (upper panel) and B1-bS (lower panel). The data are from Spitzer MIPS (24 and 70 ${\mu}$m), CSO SHARC (350 ${\mu}$m), JCMT SCUBA (850 ${\mu}$m), SMA (1.1 mm and 1.3 mm), NMA (3.3 mm), and VLA (7 mm). Arrows mark the upper limits. The solid red  curves are the single-component greybody fits with T$_{\rm dust}$ ${\thickapprox}$ 16 K and ${\beta}$ = 2.0 for B1-bN and with T$_{\rm dust}$ ${\thickapprox}$ 18 K and ${\beta}$ = 1.3 for B1-bS. 
The dashed blue curves are the results of the two component model with a compact blackbody component and an extended greybody component.
The parameters of the two component models are T$_{\rm dust}$(ext) ${\thickapprox}$ 11 K, ${\beta}$(ext) = 2.0, and T$_{\rm dust}$(cmp) = 23 K for B1-bN, and T$_{\rm dust}$(ext) ${\thickapprox}$ 16 K, ${\beta}$(ext) = 1.5, and T$_{\rm dust}$(cmp) ${\thickapprox}$ 24 K for B1-bS.
\label{fig7}}
\end{figure}
\clearpage

\begin{deluxetable}{cllll}
\tabletypesize{\small}
\renewcommand{\arraystretch}{0.65} 
\tablecolumns{5}
\tablewidth{0pc}
\tablecaption{The parameters of the SED fit}
\tablehead{ \colhead{Object Name}
 & \colhead{T$_{\rm dust}$}   & \colhead{${\beta}$} & \colhead{$\tau_{\rm 230 GHz}$} & \colhead{$r$} \\
 & \colhead{(K)} & \colhead{ } & \colhead{} & \colhead{(\arcsec)} }
\startdata
B1-bN & 15.7${\pm}$1.3	&	2.0${\pm}$0.4	& 0.15${\pm}$0.05 & 0.9${\pm}$0.2	 \\
B1-bS & 18.3${\pm}$2.1	&	1.3${\pm}$0.2  & 0.09${\pm}$0.08 & 	1.2${\pm}$1.2
\enddata
\end{deluxetable}

\begin{deluxetable}{clllc}
\tabletypesize{\small}
\renewcommand{\arraystretch}{0.65} 
\tablecolumns{5}
\tablewidth{0pc}
\tablecaption{Physical Properties of B1-bN and B1-bS}
\tablehead{ \colhead{Object Name}
 &  \colhead{L$_{\rm bol}$} & \colhead{M$_{\rm env}$} & \colhead{T$_{\rm bol}$} & \colhead{L$_{\rm sub}$/L$_{\rm bol}$} \\
 & \colhead{(L$_{\sun}$)} &\colhead{(M$_{\sun}$)} &\colhead{(K)} &}
\startdata
B1-bN &  	0.15 &	0.36	&	17.2	&	0.21 \\
B1-bS &  	0.31 	&	0.36	&	21.7 	& 	0.11
\enddata
\end{deluxetable}
\clearpage

\begin{figure}
\plotone{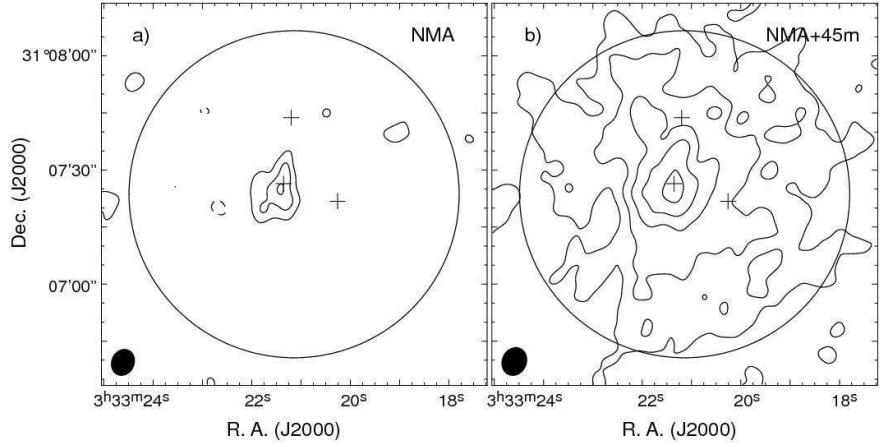}
\caption{a) Integrated intensity of the H$^{13}$CO$^+$ $J$=1--0 observed with the NMA. The integrated velocity range is from $V_{\rm LSR}$ = 5.75 to 8.25 km s$^{-1}$. Contours are drawn every 0.19 Jy beam$^{-1}$ km s$^{-1}$ (1.5 ${\sigma}$) with the lowest contour at  0.38 Jy beam$^{-1}$ km s$^{-1}$ (3 ${\sigma}$). The crosses mark the positions of B1-bN, B1-bS. and B1-bW. The circle indicates the half-power primary beam of the NMA. b) Integrated H$^{13}$CO$^+$ $J$=1--0 map with the short spacing information obtained with the NRO 45 m telescope. Contours are drawn every 0.33 Jy beam$^{-1}$ km s$^{-1}$ (3 ${\sigma}$ )with the lowest contour level of 0.33 Jy beam$^{-1}$ km s$^{-1}$ (3 ${\sigma}$).}
\label{fig8}
\end{figure}

\clearpage

\begin{figure}
\plotone{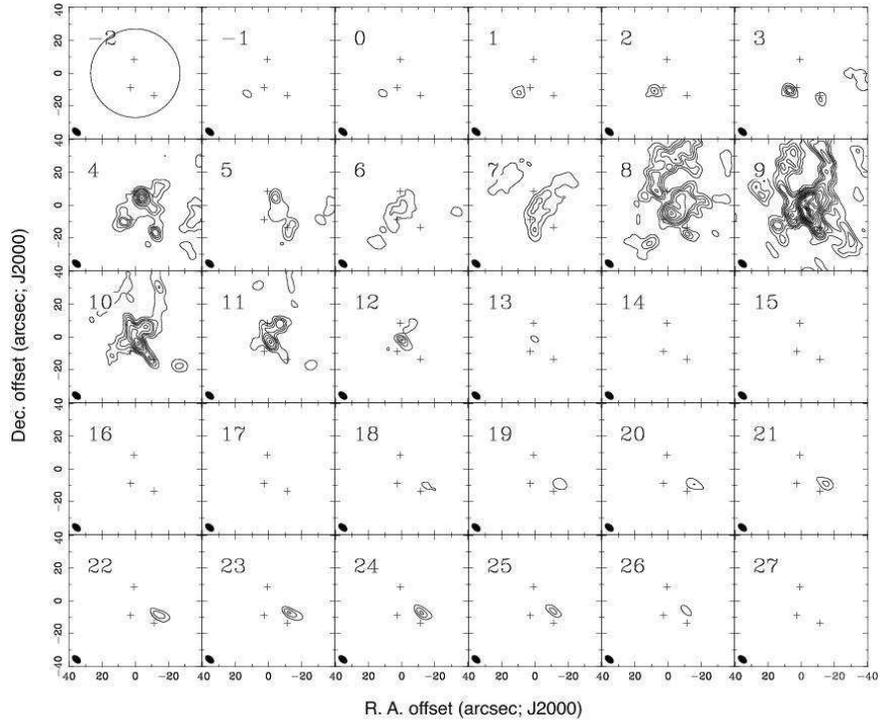}
\caption{Velocity-channel maps of the CO $J$=2--1 observed with the SMA. The central velocity of each channel is shown in the panel. The contours start at 5$\sigma$ and are drawn every 5$\sigma$ step until 20$\sigma$, and every 10$\sigma$ in the range above 20$\sigma$. The 1 $\sigma$ level is 0.1 Jy beam$^{-1}$. The crosses mark the positions of B1-bN, B1-bS. and B1-bW. The circle in the top left panel indicates the half-power primary beam of the SMA. }
\label{fig9}
\end{figure}

\clearpage

\begin{figure}
\epsscale{1}
\plotone{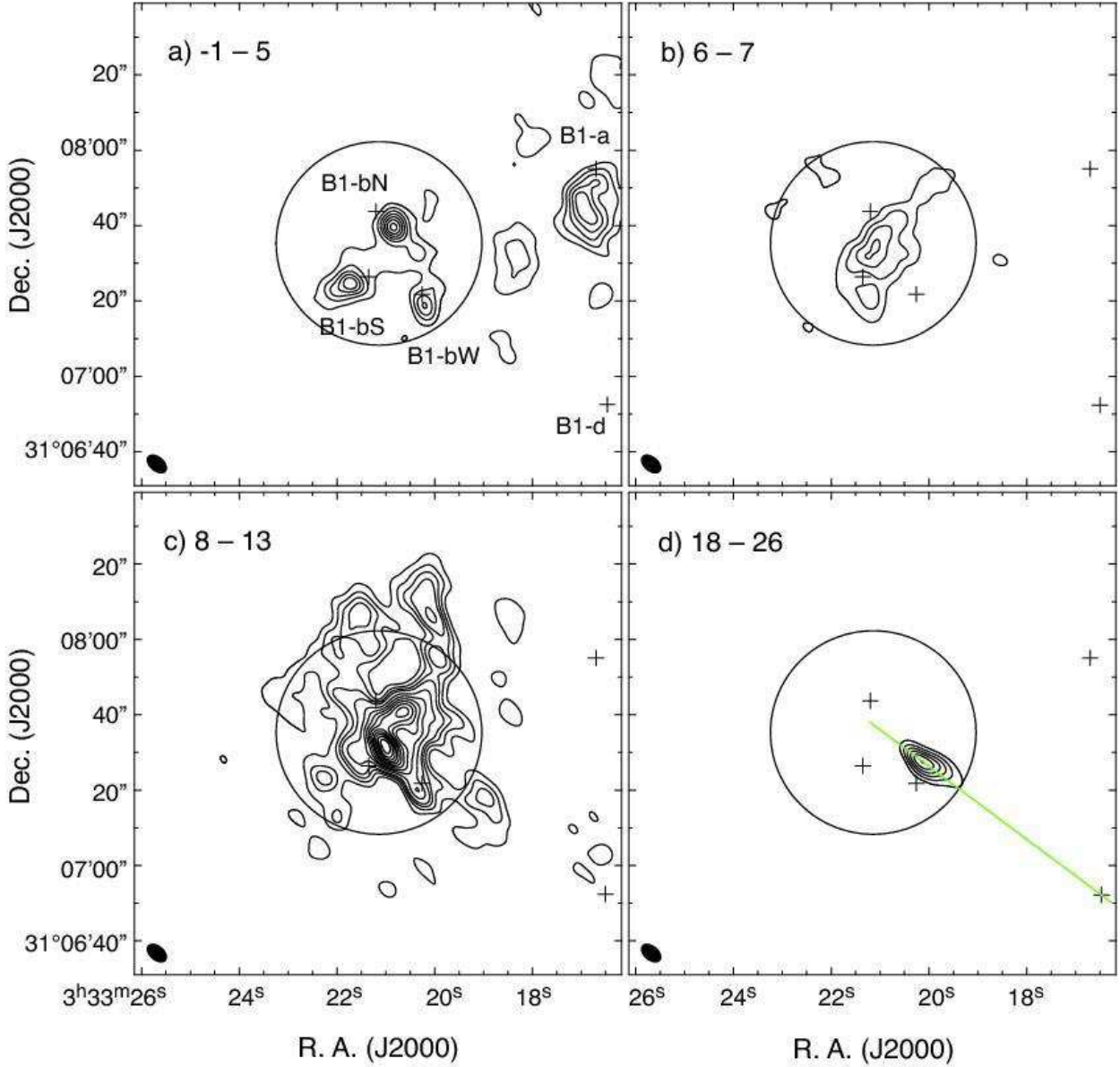}
\caption{{$^{12}$CO 2-1 emission from  B1-b region in four velocity ranges. a) $V_{\rm LSR}$ = $-$1--5 km s$^{-1}$ (blue shifted), b) $V_{\rm LSR}$ = 6--7 km s$^{-1}$ (systemic velocity), c) $V_{\rm LSR}$ = 8--13 km s$^{-1}$ (redshifted), and d) $V_{\rm LSR}$ = 18--27 km s$^{-1}$ (high-velocity red). Contours in the panels a), b), and d) are drawn every 5${\sigma}$ step with the lowest contours at 5${\sigma}$. The contours in panel c) start at 5$\sigma$ and are drawn every 5$\sigma$ step until 20$\sigma$, and every 10$\sigma$ in the range above 20$\sigma$. The 1$\sigma$ levels are 0.26 Jy beam$^{-1}$ km s$^{-1}$ in panel a), 0.14 Jy beam$^{-1}$ km s$^{-1}$ in panel b), 0.24 Jy beam$^{-1}$ km s$^{-1}$ in panel c), and 0.3 Jy beam$^{-1}$ km s$^{-1}$ in panel d). The crosses indicate the positions of B1-bN, B1-bS,  B1-bW, B1-a, and B1-d.  The open circle indicates the SMA field of view at 230 GHz, and the filled ellipse indicates the synthesized beam. The green line in the panel d) indicates the cut of the P-V diagram shown in Figure 12. Note that the primary beam correction is not applied to the data.}
\label{fig10}}
\end{figure}

\clearpage

\begin{figure}
\epsscale{1}
\plotone{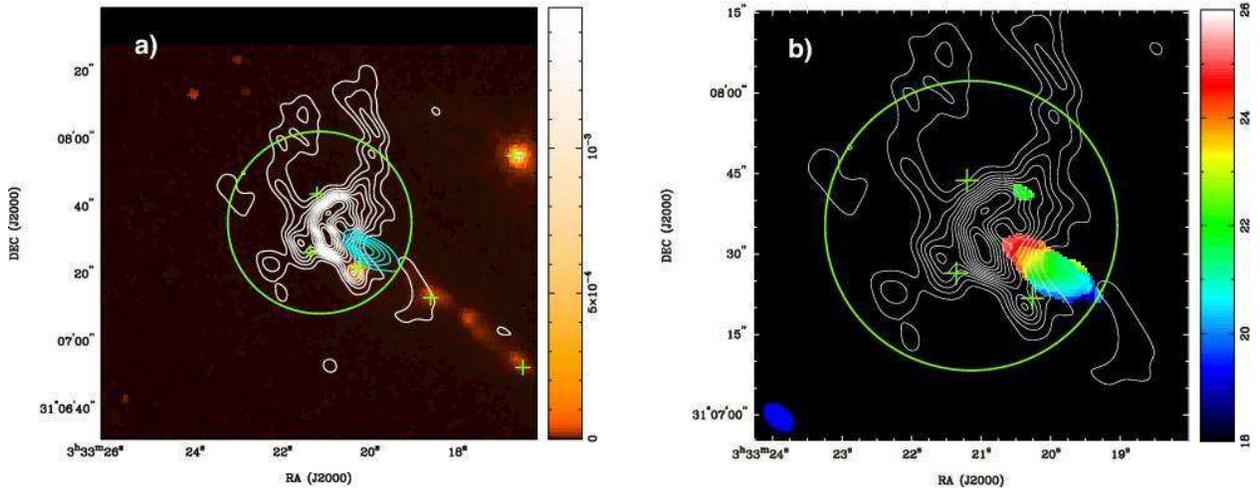}
\caption{{a) $^{12}$CO 2-1 emission at $V_{\rm LSR}$ = 9 km s$^{-1}$ (white contours, 10$\sigma$ step with the 1$\sigma$ level of 0.1 Jy beam$^{-1}$) and the high-velocity redshifted component (blue contours, 5 $\sigma$ step with the 1$\sigma$ level of 0.3 Jy beam$^{-1}$ km s$^{-1}$) on top of the {\it Spitzer} IRAC band 2 image. The crosses are the locations of B1-bN, B1-bS, B1-bW, B1-a, B1-d, and MH5.  b) Moment 1 map of the high-velocity redshifted component (color) overlaid on the  $^{12}$CO 2-1 emission at $V_{\rm LSR}$ = 9 km s$^{-1}$ (white contours). The contour levels are the same as those of panel a).}
  \label{fig11}}
\end{figure}
\clearpage

\begin{figure}
\plotone{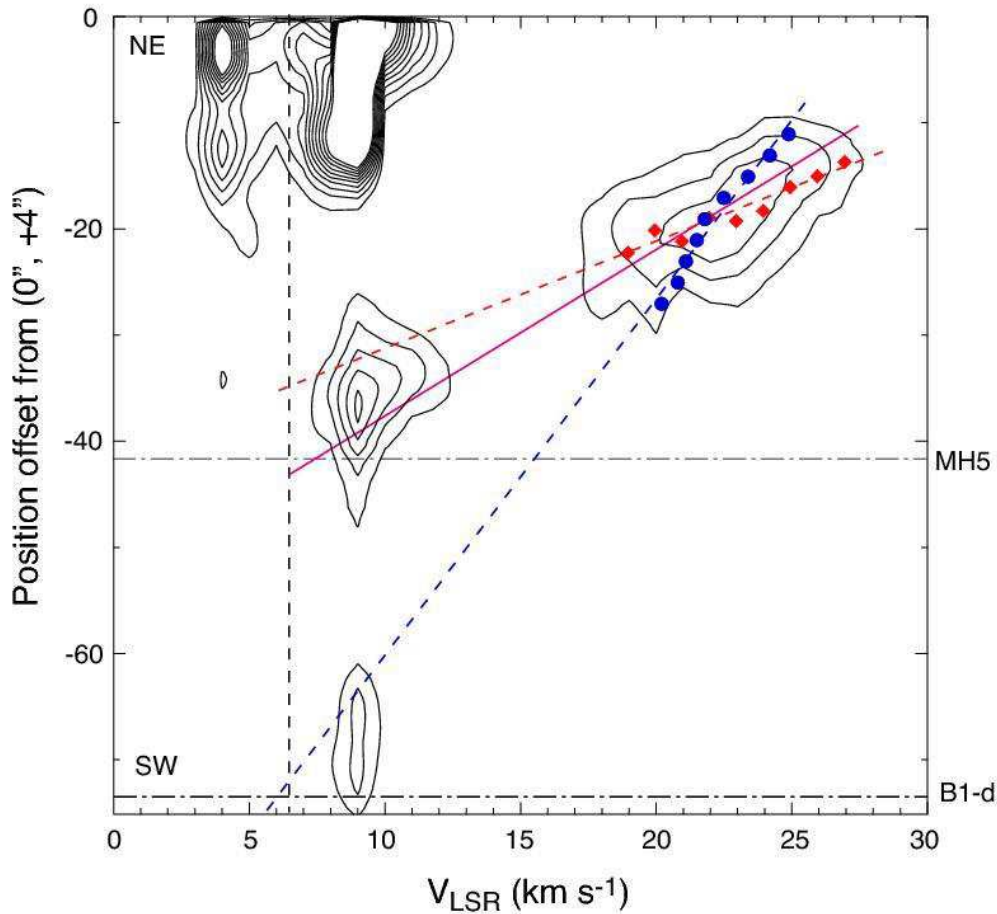}
\caption{P-V diagram of the CO $J$=2--1 along the NE-SW cut through the high-velocity CO knot and B1-d (P.A. = 53$^{\circ}$, green line in Figure 10d). Contours are drawn every 0.3 Jy beam$^{-1}$ with the lowest contour at 0.3 Jy beam$^{-1}$. {The magenta line delineates the major axis of the high-velocity emission component in the P-V diagram determined by the two dimensional gaussian fitting.
The red dots are the intensity weighted positions of the CO at different velocities. 
The blue dots are the intensity-weighted mean velocities of the CO line at the positions along the cut. The red and blue dashed lines are the linear fits of the intensity-weighted positions and velocities, respectively} The vertical dashed line labels the systemic velocity of the cloud, 6.5 km s$^{-1}$. The horizontal dash-dotted lines denote the locations of MH5 and B1-d.}
\label{fig12}
\end{figure}
\clearpage

\begin{figure}
\plotone{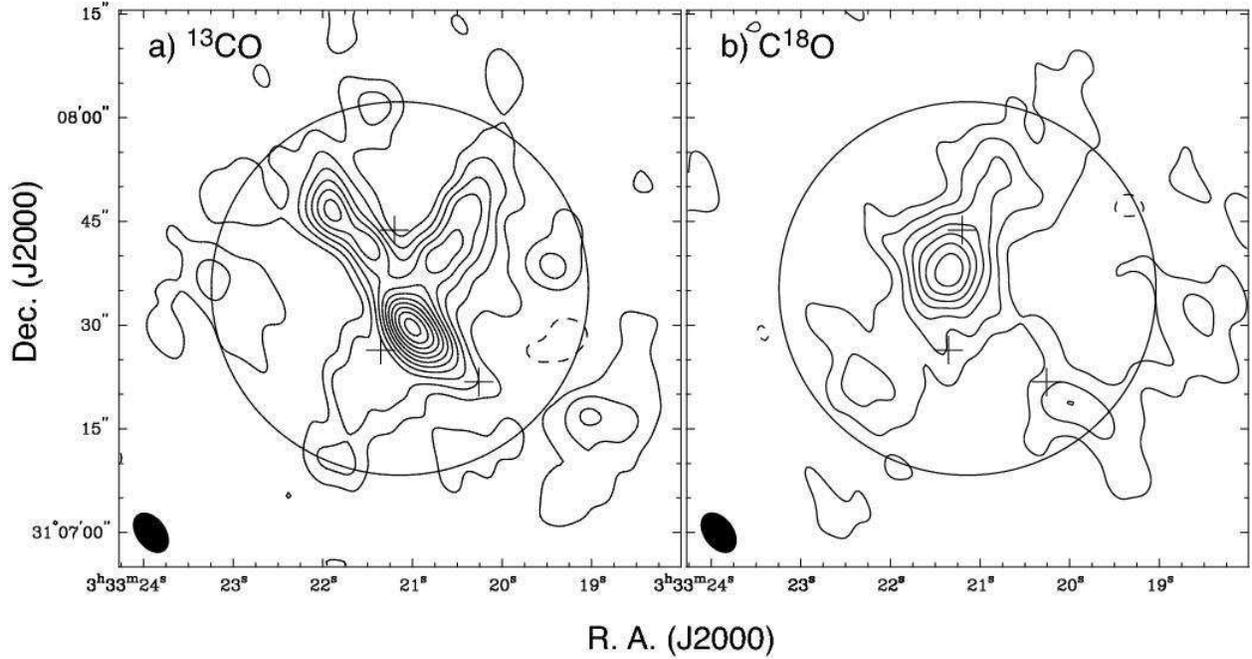}
\caption{a): $^{13}$CO $J$=2--1 emission integrated over the velocity range from $V_{\rm LSR}$ = 5 km s$^{-1}$ to 11 km s$^{-1}$. Contour levels are from 3 ${\sigma}$ by steps of 3 ${\sigma}$, with 1 ${\sigma}$ = 0.13 Jy beam$^{-1}$ km s$^{-1}$. 
b): Integrated intensity map of the C$^{18}$O $J$=2--1emission. The integrated velocity range is from $V_{\rm LSR}$ = 5.75 to 8.25 km s$^{=1}$.Contours are drawn every 3 ${\sigma}$ with 1 ${\sigma}$ = 0.095 Jy beam$^{-1}$ km s$^{-1}$.  The crosses indicate the positions of B1-bN, B1-bS, and B1-bW. The circle indicates the half-power primary beam of the SMA.}
\label{fig13}
\end{figure}
\clearpage

\begin{figure}
\plotone{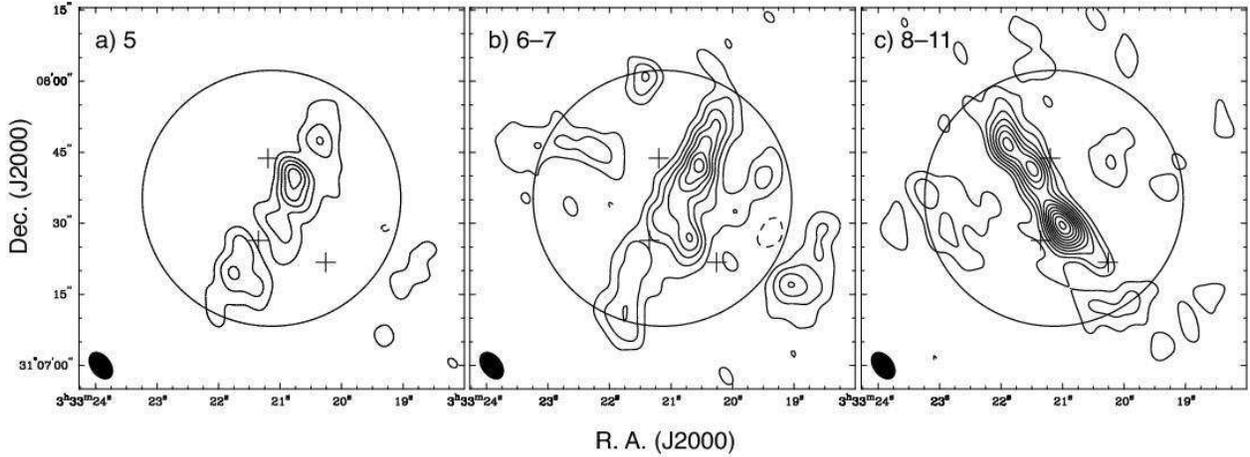}
\caption{$^{13}$CO $J$=2--1 emission in three velocity ranges. a) $V_{\rm LSR}$ = 5 km s$^{-1}$, b) $V_{\rm LSR}$ = 6--7 km s$^{-1}$, and c) $V_{\rm LSR}$ = 8--11 km s$^{-1}$. Contour levels are from 3 ${\sigma}$ by steps of 3 ${\sigma}$. The 1 ${\sigma}$ levels are 0.05 Jy beam$^{-1}$ km s$^{-1}$ in panel a), 0.07 Jy beam$^{-1}$ km s$^{-1}$ in panel b). and 0.1 Jy beam$^{-1}$ km s$^{-1}$ in panel c). 
The crosses indicate the positions of B1-bN, B1-bS, and B1-bW. The circle indicates the half-power primary beam of the SMA.}
\label{fig14}
\end{figure}
\clearpage

\begin{deluxetable}{lcccccc}
\renewcommand{\arraystretch}{0.75} 
\tabletypesize{\small}
\tablecolumns{7}
\tablewidth{0pc}
\tablecaption{Outflow Parameters$^{\rm a}$}
\tablehead{ \colhead{Component}  & \colhead{$M$}  &\colhead{$P$} &\colhead{$V_{\rm flow}$} &\colhead{$R$}  &\colhead{$t_d$}  &\colhead{$F$}\\
\colhead{} & \colhead{($M_{\sun}$) }	& \colhead{($M_{\sun}$ km s$^{-1}$)} 	&\colhead{(km s$^{-1}$)} & \colhead{(AU)} 	&\colhead{(yr)} 	&\colhead{($M_{\sun}$ km s$^{-1}$ yr$^{-1}$)} }
\startdata
\sidehead{B1-bN$^{\rm b}$}
Blue & 5.5$\times$10$^{-4}$ & 1.3$\times$10$^{-3}$	& 2.4 &1400 &	2800 	&	4.7$\times$10$^{-7}$		\\
Red$^{\rm c}$ & 2.4$\times$10$^{-4}$	& 5.0$\times$10$^{-4}$	& 2.1 &	1700	&	4000 &	1.3$\times$10$^{-7}$ \\
\sidehead{B1-bS$^{\rm d}$}
Blue$^{\rm c}$ & 9.6$\times$10$^{-5}$	& 4.0$\times$10$^{-4}$	& 4.1 &	1600 &	1800	&	2.2$\times$10$^{-7}$ \\
Red & 2.2$\times$10$^{-3}$ & 5.5$\times$10$^{-3}$	& 2.5 &	1600 &	3000 &	1.9$\times$10$^{-6}$\\
Extended red & 5.1$\times$10$^{-3}$ & 1.1$\times$10$^{-2}$ & 2.2 &  9200 & 20000 & 5.8$\times$10$^{-7}$\\
\sidehead{B1-bW$^{\rm e}$}
Blue$^{\rm c}$ & 6.1$\times$10$^{-5}$ & 1.2$\times$10$^{-4}$ & 1.9 & 700 & 1700 & 7.0$\times$10$^{-8}$ \\
Red$^{\rm c}$ & 1.0$\times$10$^{-4}$ & 3.2$\times$10$^{-4}$ & 3.0 & 600 & 900 & 3.6$\times$10$^{-7}$\\
\sidehead{High-velocity component$^{\rm f}$}
Red$^{\rm c,g}$ & 1.3$\times$10$^{-4}$ & 2.0$\times$10$^{-3}$ & 15.3 & 5500 & 1700 & 1.1$\times$10$^{-6}$ \\
Red$^{\rm c,h}$ &  & &  & 14300 & 4500 & 4.4$\times$10$^{-7}$ \\
\enddata
\tablenotetext{a}{All the measured values are already corrected for primary beam correction.} 
\tablenotetext{b}{$V_{\rm sys}$ = 7.2 km s$^{-1}$ derived from the N$_2$D$^+$ $J$=3--2 line \citep{Huang13}.}
\tablenotetext{c}{No opacity correction was applied because there was no contour part in $^{13}$CO line emission.}
\tablenotetext{d}{$V_{\rm sys}$ = 6.3 km s$^{-1}$ derived from the N$_2$D$^+$ $J$=3--2 line \citep{Huang13}.}
\tablenotetext{e}{$V_{\rm sys}$ = 6.1 km s$^{-1}$ derived from the C$^{18}$O $J$=2--1 data.}
\tablenotetext{f}{$V_{\rm sys}$ = 6.5 km s$^{-1}$ derived from the H$^{13}$CO$^+$ spectrum.}
\tablenotetext{g}{{ The driving source is assumed to be MH5.}}
\tablenotetext{h}{The driving source is assumed to be B1-d.}

\end{deluxetable}
\clearpage

\end{document}